\documentclass[a4paper,11pt]{article}

\usepackage{jcappub}
\usepackage{amsmath}
\usepackage{graphicx}
\usepackage{latexsym}
\usepackage{xspace}
\usepackage{color}
\usepackage{hyperref} 
\usepackage{bm}
\usepackage{relsize}
\usepackage{mathdots}
\usepackage{makecell}
\usepackage[caption=false]{subfig}

\makeatletter
\gdef\@fpheader{}
\makeatother

\newcommand{\ie}{\textit{i.e.}\xspace}

\DeclareMathOperator{\sign}{sign}
\newcommand{\Pade}{Pad\'e }

\newcommand{\dd}{\mathrm{d}}
\newcommand{\ee}{e}

\newcommand{\sss}[1]{{\scriptscriptstyle{#1}}}

\newcommand{\uPl}{\mathrm{Pl}}
\newcommand{\uin}{\mathrm{in}}

\newcommand{\uend}{\mathrm{end}}

\newcommand{\uS}{\mathrm{S}}

\newcommand{\usssS}{\sss{\uS}}

\newcommand{\usssPl}{\sss{\uPl}}

\newcommand{\nS}{n_\usssS}

\newcommand{\calP}{\mathcal{P}}

\newcommand{\Mp}{M_\usssPl}

\newcommand{\efolds}{$e$-folds~}

\newcommand{\beq}{\begin{equation}}
\newcommand{\eeq}{\end{equation}}
\newcommand{\bea}{\begin{eqnarray}}
\newcommand{\eea}{\end{eqnarray}}

\newlength{\wsingfig}
\newlength{\wdblefig}
\newlength{\wquadfig}
\newlength{\wtriplefig}
\newcommand{\Eq}[1]{Eq.~(\ref{#1})}
\newcommand{\Eqs}[1]{Eqs.~(\ref{#1})}
\newcommand{\Fig}[1]{Fig.~{\ref{#1}}}

\newcommand{\Ref}[1]{Ref.~{\cite{#1}}}
\newcommand{\Refs}[1]{Refs.~{\cite{#1}}}

\newcommand{\widthdouble}{7.5cm}

\subheader{}

\title{The Hubble Flow of Plateau Inflation}

\author[a,b]{Dries Coone,}

\author[a]{Diederik Roest}

\author[c]{and Vincent Vennin}

\affiliation[a]{Van Swinderen Institute for Particle Physics and Gravity, University of Groningen,
Nijenborgh 4, 9747 AG Groningen, (The Netherlands)}

\affiliation[b]{Theoretische Natuurkunde, Vrije Universiteit Brussel and The International Solvay Institutes Pleinlaan 2, B-1050 Brussels, (Belgium)}

\affiliation[c]{Institute of Cosmology \& Gravitation, University of Portsmouth, Dennis Sciama Building, Burnaby Road, Portsmouth, PO1 3FX, (United Kingdom)}

\emailAdd{a.a.coone@rug.nl}
\emailAdd{d.roest@rug.nl}
\emailAdd{vincent.vennin@port.ac.uk}

\date{today}

\begin{document}

\abstract{In the absence of CMB precision measurements, a Taylor expansion has often been invoked to parametrize the Hubble flow function during inflation. The standard ``horizon flow'' procedure implicitly relies on this assumption. However, the recent Planck results indicate a strong preference for plateau inflation, which suggests the use of \Pade approximants instead. We propose a novel method that provides analytic solutions of the flow equations for a given parametrization of the Hubble function. This method is illustrated in the Taylor and \Pade cases, for low order expansions. We then present the results of a full numerical treatment scanning larger order expansions, and compare these parametrizations in terms of convergence, prior dependence, predictivity and compatibility with the data. Finally, we highlight the implications for potential reconstruction methods.}


\arxivnumber{1507.00096}

\maketitle
\section{Introduction}
\label{sec:intro}
Inflation is one of the leading paradigms for explaining the physical conditions that prevailed in the very early Universe~\cite{Starobinsky:1980te,Sato:1980yn,Guth:1980zm,Linde:1981mu,Albrecht:1982wi,Linde:1983gd}. It consists in a phase of accelerated expansion that solves the standard hot Big Bang model problems, and provides a causal mechanism for generating inhomogeneities on cosmological scales~\cite{Starobinsky:1979ty,Mukhanov:1981xt,Hawking:1982cz,Starobinsky:1982ee,Guth:1982ec,Bardeen:1983qw}. 
The most recent Planck measurements~\cite{Adam:2015rua,Ade:2015lrj,Ade:2015ava} of the Cosmic Microwave Background (CMB) indicate that these cosmological perturbations are almost scale invariant, with undetected level of non-Gaussianities and isocurvature components. At this stage, the full set of observations can therefore be accounted for in the minimal setup, where inflation is driven by a single scalar inflaton field $\phi$ with canonical kinetic term, minimally coupled to gravity and evolving in some potential $V(\phi)$.

However, since the inflationary mechanism is supposed to take place at very high energies, in a regime where particle physics is not known and has not been tested in accelerators, the physical nature of the inflaton and its relation with the standard model of particle physics and its extensions remain elusive. The only condition on $V$ is that it should be sufficiently flat to support inflation, but otherwise the multitude of inflaton candidates (with associated potentials) makes the theory as a whole hardly tractable, unless one restricts to a specific model or scan them all one by one~\cite{Martin:2014vha,Martin:2013nzq,Martin:2014nya,Martin:2014rqa}.

Another strategy consists in developing model independent approaches and in studying generic parametrizations of the inflationary dynamics (see, for instance, \Refs{Barrow:1990vx,Alcaniz:2006nu,Barrow:2007zr,Mukhanov:2013tua, Roest:2013fha, Boubekeur:2014xva}). Given the large volume in parameter space left by the first CMB anisotropy measurements~\cite{Hanany:2000qf,Lange:2000iq,Balbi:2000tg}, Hoffman, Turner and Kinney~\cite{Hoffman:2000ue, Kinney:2002qn} proposed to study ``generic'' Hubble flow dynamics by performing a Monte Carlo analysis over a set of Hubble functions. This procedure, dubbed ``horizon flow'', was shown to yield the ``typical'' predictions $r\simeq 0$ or $r\simeq 16 (1-\nS)/3$, where $\nS$ is the scalar spectral index and $r$ is the tensor-to-scalar ratio. When $r\simeq 0$, it was found that generically, $\nS<0.85$ or $\nS>1$. However, Liddle pointed out~\cite{Liddle:2003py} that this parametrization implicitly relies on Taylor expanding the Hubble function to some order, implying some specific potentials. In \Refs{Chongchitnan:2005pf,Ramirez:2005cy,Vennin:2014xta}, a more detailed analysis of the problem revealed that the phenomenological class of inflationary potentials sampled by the procedure is indeed responsible for these predictions.
In particular, it should not come as a surprise that potential reconstruction methods based on the horizon flow procedure naturally give rise to chaotic-like potentials, despite the data. In some sense, those are already oversampled in the prior of the method.  

Phenomenological parametrizations of inflation always contain some bias towards a specific class of dynamics and it is therefore often tricky to blindly use them in order to obtain physical information from, say, observational data. This is why in this paper, we reverse the problem and ask the following question. Given what we observationally know, what is the most sensible parametrization of inflation that incorporates current observational constraints? What prescriptions should be used when studying other aspects of the early Universe, where one needs to model inflationary dynamics phenomenologically?

The Planck satellite results have attracted attention to a particular set of inflationary scenarios~\cite{Guth:2013sya,Martin:2013nzq}, plateau inflation. One possibility to include this piece of information in the way we parametrize inflation is to make use of \Pade approximants of the Hubble function instead of Taylor expansions. The rest of this paper is therefore organized as follows. In section~\ref{sec:FlowDynamics}, we set the main notations and propose a new method to analytically integrate the flow equations. This relies on making use of the field redefinition invariance of the problem, in order to identify integration constants \textit{a la} Noether. In section~\ref{sec:analytics}, we apply this method at low order to Taylor and \Pade expansions of the Hubble function, and compare both results. (In appendix~\ref{sec:InverseTaylor}, we also give the results for an inverse Taylor expansion, in order to further illustrate how our method works in practice.) Then, in section~\ref{sec:numeric}, we provide a numerical analysis of these two parametrizations, when generalized to higher orders. In particular, we compare the way they sample inflationary parameter space as a function of the truncation order, discuss their prior dependence and comment on their agreement (or lack thereof) with observations. Finally, in section~\ref{sec:Conclusion}, we recap our main results and draw a few conclusions, including comments on potential reconstruction with this method.
\section{Hubble Flow Dynamics}
\label{sec:FlowDynamics}
The horizon flow formalism relies on the introduction of a set of ``flow parameters'' characterizing the way the Hubble scale evolves in time. There are several possible sets of such parameters. For example, let us consider the so-called ``Hubble flow parameters'', defined by the flow equations \cite{Schwarz:2001vv,Schwarz:2004tz} 
\beq
\label{eq:epsFlow}
\epsilon_{i+1}=\frac{\dd\ln\epsilon_i}{\dd N}\, .
\eeq
The hierarchy is started at $\epsilon_0\equiv H_\uin/H$, where $N\equiv \ln a$ is the number of \efolds and increases as time proceeds. Since $\epsilon_1=-\dot{H}/H^2=1-\ddot{a}/(aH^2)$, inflation ($\ddot{a}>0$) takes place provided $\epsilon_1<1$. These parameters can be written in terms of the Hubble function $H(\phi)$ and its derivatives. For the lowest orders, one has 
\begin{equation}
\begin{aligned}
\epsilon_1 &=2 \left( \frac{H^\prime}{H} \right)^2 \, , \quad\quad
\epsilon_2=4 \left[\left(\frac{H^\prime}{H}\right)^2-\frac{H^{\prime\prime}}{H}\right]\, ,&\\
\epsilon_3 &=2 \left[2\left(\frac{H^\prime}{H}\right)^2+\frac{H^{\prime\prime\prime}}{H^\prime}-3\frac{H^ {\prime\prime}}{H}\right] \left(1-\frac{HH^{\prime\prime}}{{H^\prime}^2}\right)^{-1}\,. &
\end{aligned}
\label{eq:eps:H}
\end{equation}
Throughout this paper we set the reduced Planck mass $\Mp =1$, dots mean derivating with respect to cosmic time and primes refer to derivating with respect to the inflaton field $\phi$.

The horizon flow strategy rests on solving a truncated hierarchy of flow equations for a given set of flow parameters. Since these flow parameters can always be written in terms of the $H(\phi)$ function and its derivatives, as in \Eqs{eq:eps:H}, this procedure thus relies on the assumption that some combination of $H$ and its derivatives, corresponding to the first vanishing flow parameter, is zero. Interpreted as a differential equation for the $H(\phi)$ function, this means that $H(\phi)$ is parametrized in a certain manner, involving a finite number of constant free parameters.

For example, if the Hubble flow hierarchy is truncated at some order $M$, \ie if one assumes $\epsilon_l=0$ for $l>M$, then $\epsilon_M$ is constant, and $H \propto \exp(a_1\exp(a_2\cdots\exp(a_M N)\cdots))$ where the exponential function is composed $M$ times. As another example, if one makes use of the ${}^l\lambda$ parameters, widely used in the horizon flow literature and defined as~\cite{Kinney:2002qn} 
\beq
\label{eq:lambda:def}
{}^l\lambda_H=2\frac{\left(H^\prime\right)^{l-1}}{H^l}\frac{\dd^{l+1}H}{\dd\phi^{l+1}} \,,
\eeq
truncating the hierarchy at order $M$ means that $\dd^{M+1}H/\dd\phi^{M+1}=0$, hence $H(\phi)$ has a polynomial form of degree $M$. In general, one can see that truncating a specific flow hierarchy always boils down to parametrizing the Hubble function in a specific manner.

Conversely, to any parametrization of the Hubble function, one can associate a specific dynamical system. Let $H\left(\phi,a_1,a_2,\ldots,a_n\right)$ be a given parametrization, where the coefficients $a_i$ stem from some (e.g.~Taylor or \Pade) expansion truncated at some order $n$. The $n+1$ first derivatives of this function with respect to the inflaton field $\phi$ can be calculated, and one can invert the system to extract the $n+2$ variables $\left\lbrace \phi, a_1,a_2,\ldots,a_n, H^{(n+1)}  \right\rbrace$ in terms of $\left\lbrace H, H', H'',\ldots, H^{(n)} \right\rbrace$. Of particular interest is the last entry of this solution, which relates the $(n+1)^\mathrm{th}$ derivative of the Hubble function $H^{(n+1)}$ to the lower order ones. In terms of the Hubble flow hierarchy, this means that $\epsilon_{n+1}$ can be expressed in terms of the $n$ first Hubble flow parameters only. The flow equations~(\ref{eq:epsFlow}), for $1\leq i \leq n$, thus form a closed dynamical system. It is important to stress that all physical input resides in this truncation: how $\epsilon_{n+1}$ is expressed in terms of all preceding flow parameters fully determines the dynamical system and hence the inflationary predictions. 
 
Moreover, the flow dynamics is insensitive to the actual value of the inflaton field $\phi$ and hence the transformation $\phi \rightarrow \phi + \delta \phi$ leaves this system invariant. For the expansions that we consider, indeed, the functional form of $H(\phi)$ does not change under this shift, implying that there is a degeneracy in the parameters. Amongst the $n$ parameters, one combination can therefore be absorbed by the shift transformation, while the remaining $n-1$ combinations are invariant. The later are therefore constants of motion, and the space of inflationary solutions has dimension $n-1$. 

In the following section, we show how these constants of motion can be derived in practice, and we apply this method to second order Hubble and \Pade expansions of the Hubble function. In these cases, a single constant of motion will be obtained. Therefore, at fixed number of \efolds $\Delta N_*$ between the Hubble exit time of the CMB pivot scale and the end of inflation, a one-to-one relation between $\nS$ and $r$ will be obtained.

A last remark is in order. Though the systems will be solved exactly and independently of the slow-roll approximation, in practice, the scalar spectral index and the tensor-to-scalar ratio will be calculated from the flow parameters at Hubble exit time thanks to the relations~\cite{Gong:2001he,Martin:2013uma}
\beq  \label{eq:nsr:srnlo}
\nS =1-2\epsilon_{1*}-\epsilon_{2*}-2\epsilon_{1*}^2 -(2C+3)\epsilon_{1*}\epsilon_{2*}-C\epsilon_{2*}\epsilon_{3*}\, , \quad
r=16\epsilon_{1*}\left(1+C\epsilon_{2*}\right)\, ,
\eeq
which are valid at second order in slow roll. Here, $C\equiv\gamma_\mathrm{E}+\ln 2-2\simeq -0.7296$, $\gamma_\mathrm{E}$ being the Euler constant. In the Planck preferred regions, the above expressions are valid and can therefore be used to compare predictions with observations.
\section{Analytical Integration of the Hubble Flow}
\label{sec:analytics}
In this section, we apply the method sketched in section~\ref{sec:FlowDynamics} to two toy cases: a Taylor expansion of the Hubble function at quadratic order and a \Pade expansion at linear order. We obtain analytical expressions for the inflationary trajectories in the parameter space $(\epsilon_1,\epsilon_2)$, as well as for the number of \efolds realized along these trajectories. Finally, we display in each case the corresponding values of $\nS$ and $r$.
\subsection{Taylor Expansion}
\label{sec:analytics:exact:tay}
We first illustrate our method by considering the well-known case of a quadratic Hubble function:
\beq
 \label{H-Taylor}
H = H_0 \left( 1 + a \phi + b \phi^2 \right) \, .
\eeq
This case was already solved in \Ref{Vennin:2014xta} but employing a different approach.  Making use of \Eqs{eq:eps:H}, the two first flow parameters are given by
\beq
\label{eq:Taylor2:epsOfPhi}
	\epsilon_1 = 2\left(\frac{a+2b\phi}{1+a\phi+b\phi^2}\right)^2 \,, \quad
	\epsilon_2 = 4\left[\left(\frac{a+2b\phi}{1+a\phi+b\phi^2}\right)^2 -\frac{2b}{1+a\phi+b\phi^2}\right] \,.
\eeq
Let us derive the constant of motion. Under the inflaton shift transformation $\phi \rightarrow \phi+\delta\phi$, the functional form of \Eq{H-Taylor} remains unchanged if the coefficients of the expansion change according to
\beq
	a \rightarrow \frac{a+2 b\delta\phi}{1+a\delta\phi+b\delta\phi^2}\,, \quad
	b \to \frac{b}{1+a\delta\phi+b\delta\phi^2} \,,
\eeq
where $H_0$ has also to be rescaled according to $H_0 \rightarrow H_0 \left(1+a\delta\phi+b\delta\phi^2\right)$. If one moves to the specific gauge where $a$ vanishes, \ie if one takes $\delta\phi=-a/(2b)$, then the shifted $b$ coefficient, $b^2/(b-a^2/4)$, is gauge invariant. This implies that the following combination is invariant under the inflaton shift\footnote{The case with $a^2 = 4b$ or, equivalently, $\epsilon_1 = \epsilon_2$, is singular and needs to be treated separately. It is straightforward to show that, in this case, one simply has $\epsilon_{1*}=\epsilon_{2*}=1/(1+\Delta N_*)$.}:
\beq
\label{eq:Taylor2:gamma:def}
	\gamma =\frac{ 32 b^2}{a^2-4b}=\frac{(2\epsilon_1-\epsilon_2)^2}{\epsilon_2-\epsilon_1} \,,
\eeq
where the second equality has been obtained from \Eq{eq:Taylor2:epsOfPhi}.

\begin{figure*}[t]
\begin{center}
\includegraphics[width=\widthdouble]{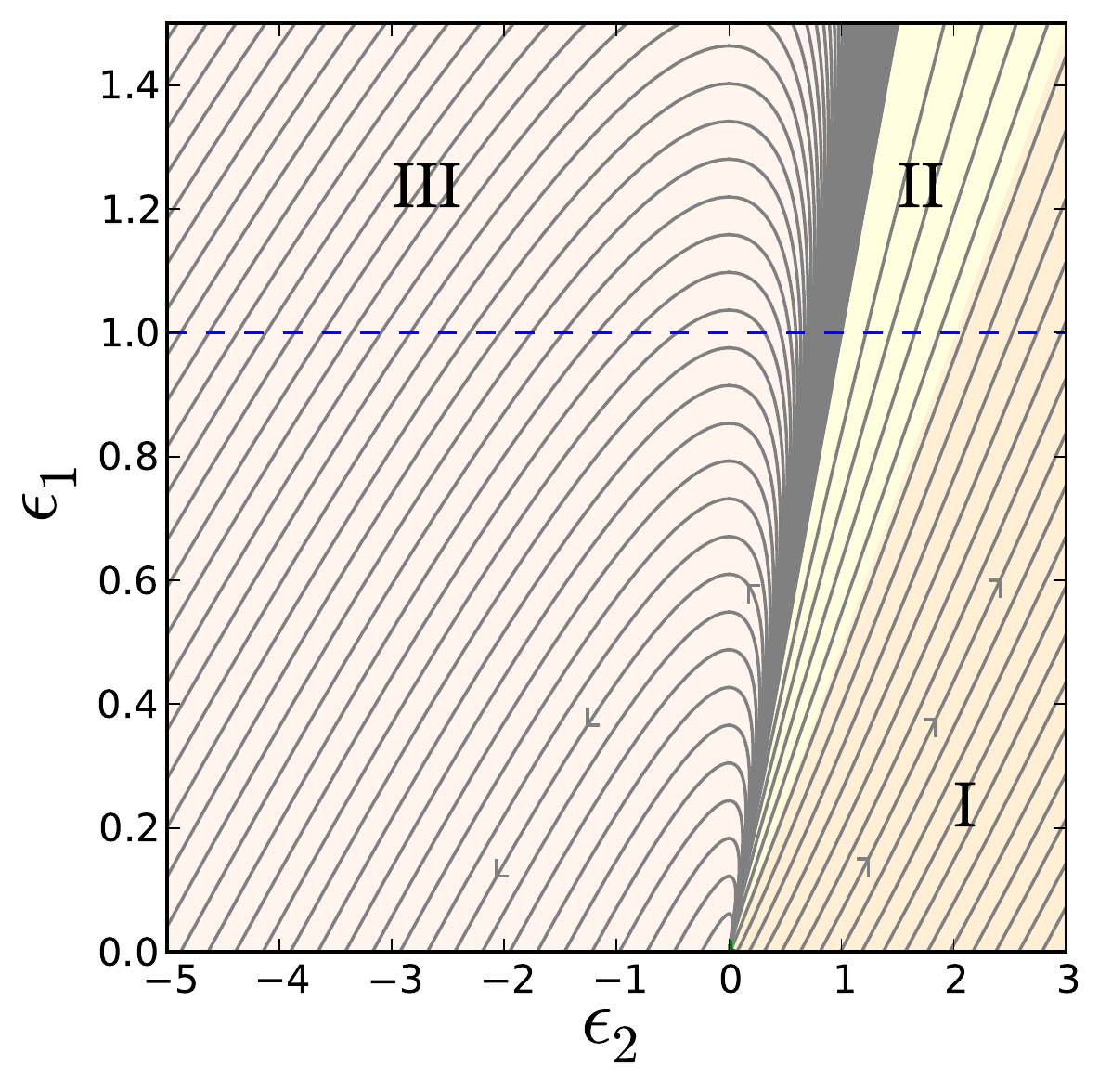}
\includegraphics[width=\widthdouble]{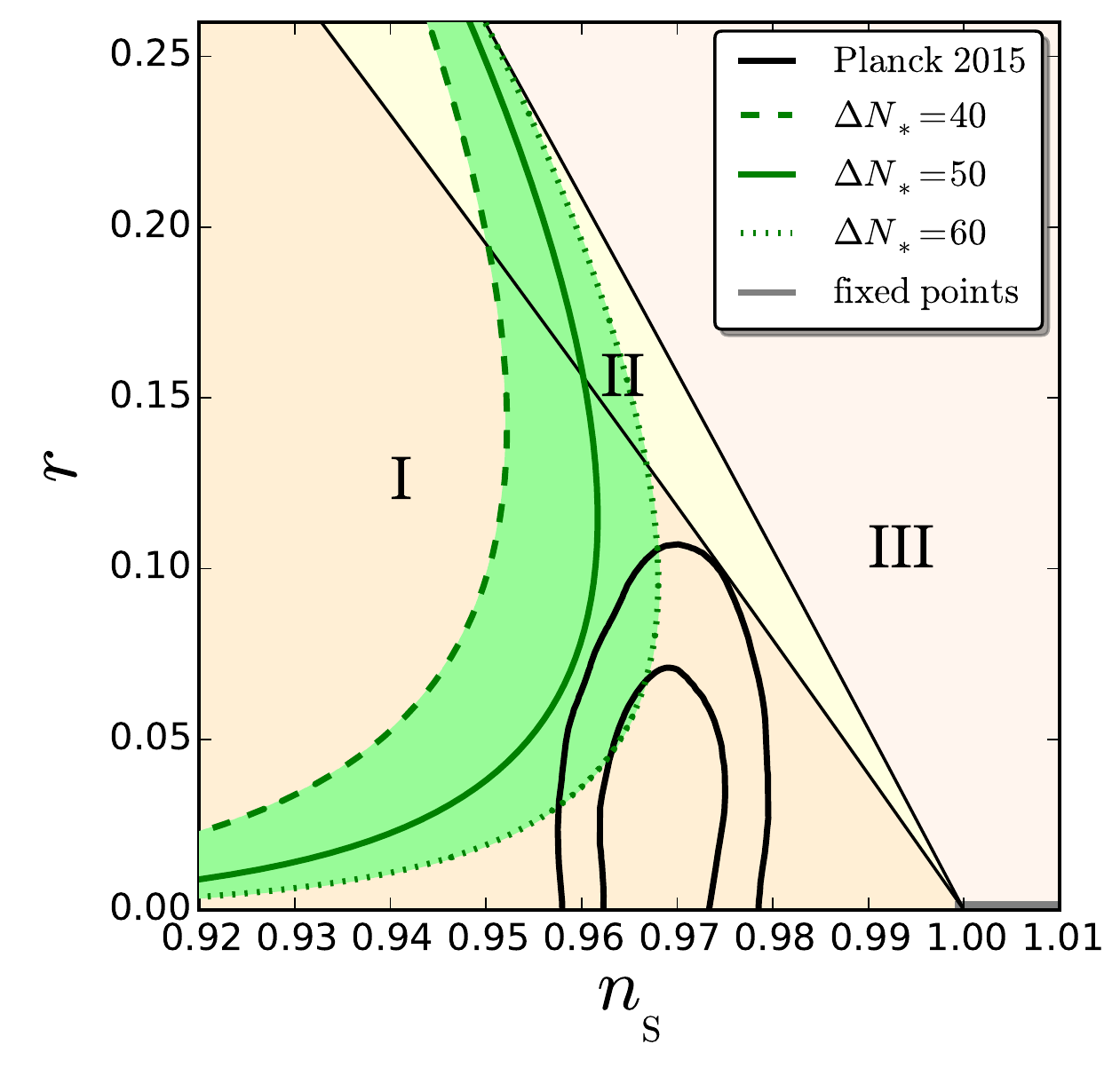}
\caption{\it Second order Taylor expansion for the Hubble function. Left panel: flow lines of the system~(\ref{eq:Taylor2:floweqs}) in the plane $(\epsilon_1,\epsilon_2)$. The arrows indicate in which direction inflation proceeds. The blue dashed line corresponds to $\epsilon_1=1$ where inflation stops. The three regions $\rm I$, $\rm II$ and $\rm III$ refer to the discussion in the main text. Right panel: Observational predictions in the $(\nS,r)$ plane, compared with the Planck 2015 $1\sigma$ and $2\sigma$ contours. The green lines stand for the values of $\nS$ and $r$ computed 40 \efolds before the end of inflation (dashed line), 50 \efolds (solid line) and 60 \efolds (dotted line). The grey segment at the bottom right stand for the fixed points $(\epsilon_1=0,\epsilon_2<0)$.}
\label{fig:Pade20}
\end{center}
\vspace{-0.5cm}
\end{figure*}

As pointed out in section~\ref{sec:FlowDynamics}, a given parametrization of the Hubble function can be translated into a specific dynamical system. For the case at hand, since \Eq{H-Taylor} implies that $H'''=0$, \Eqs{eq:eps:H} give rise to
\beq
\label{eq:eps3:Taylor}
\epsilon_3=3\epsilon_1-2\frac{\epsilon_1^2}{\epsilon_2}\, .
\eeq
This truncates the infinite set of flow equations~(\ref{eq:epsFlow}) for all the $\epsilon_i$ into a set of two differential equations for the first two flow parameters:
\beq
\label{eq:Taylor2:floweqs}
   \frac{\dd \epsilon_1}{\dd N} = \epsilon_1 \epsilon_2 \,, \qquad
   \frac{\dd \epsilon_2}{\dd N} = -2\epsilon_1^2+3\epsilon_1\epsilon_2 \,. 
\eeq
This dynamical system generates a flow through the two dimensional space $(\epsilon_1,\epsilon_2)$, which is displayed in the left panel of \Fig{fig:Pade20}. Different trajectories can be labeled by different values of the invariant parameter $\gamma$. It can easily be checked that \Eqs{eq:Taylor2:floweqs} indeed leave this particular combination invariant.

Moreover, by inverting \Eq{eq:Taylor2:gamma:def}, one can relate one of the remaining two slow-roll parameters to the other:
\begin{align}
	\label{eq:eps2:feps}
	\epsilon_2 = 2\epsilon_1+\frac{\gamma}{2}+\frac{\xi}{2}\sqrt{\gamma^2+4\gamma\epsilon_1}\, ,
\end{align}
where $\xi=\pm 1=\mathrm{sign}\left[(2\epsilon_1-\epsilon_2)\epsilon_2(\epsilon_1-\epsilon_2)\right]$ and the argument of the square root is always positive. Inserting \Eq{eq:eps2:feps} into the first of \Eqs{eq:Taylor2:floweqs} then leads to a first order differential equation for $\epsilon_1(N)$ that can be solved, and one obtains
\beq 
\label{eq:Nfeps:taylor}
\Delta N_*=N\left(\epsilon_{1,\mathrm{end}}\right)-N\left(\epsilon_{1*}\right)\, , \quad
N\left(\epsilon_1\right)= \frac{2}{\gamma+\xi \sqrt{\gamma^2+4\gamma\epsilon_1}} + \frac{\xi}{\gamma}\log\left\vert \frac{\sqrt{\gamma^2+4\gamma\epsilon_1}-\gamma} {\sqrt{\gamma^2+4\gamma\epsilon_1}+\gamma}\right\vert \,,
\eeq
where $\epsilon_{1,\mathrm{end}}=1$ (we only consider cases where inflation has a graceful exit). This expression can be inverted to yield $\epsilon_1$ as a function of $N$,
\beq
\label{eq:epsNef:taylor}
	\epsilon_1 = \frac{-\gamma W_\chi \left(-\ee^{\gamma \Delta N-1} \right)}{\left[1+W_\chi \left(-\ee^{\gamma \Delta N-1} \right)\right]^2} \,, \quad \chi = \begin{cases} 
				-1 \text{ if } \gamma(2\epsilon_1-\epsilon_2)<0 \\
				0 \text{ if } \gamma(2\epsilon_1-\epsilon_2)>0
\end{cases}\, ,
\eeq
where $\chi$ determines which branch of the Lambert function $W_\chi$ is to be used. Note that the sign of the combination $2\epsilon_1-\epsilon_2$ appearing in the definition of $\chi$ does not change during inflation, as the Hubble flow equations imply that $\dd (2\epsilon_1-\epsilon_2)/\dd N=\epsilon_1(2\epsilon_1-\epsilon_2)$. Therefore, the branch of the Lambert function does not change during inflation. One can check that the above formula matches Eqs.~(66-70) of \Ref{Vennin:2014xta} where it was first derived.

Let us now discuss the structure of the phase space diagram plotted in the left panel of \Fig{fig:Pade20}. According to the type of Hubble function one is dealing with, three possibilities must be distinguished:

\begin{minipage}{0.7\textwidth}
\begin{itemize}
\item
Firstly, if $\epsilon_2 > 2\epsilon_1$, then $\epsilon_1$ vanishes in the far past while $\epsilon_2$ takes a positive value. During inflation, both monotonically increase. This corresponds to a Hubble function that has an inverted parabolic profile whose maximum is positive (region I). In this case, $\gamma>0$ and $\xi=+1$.\\
\end{itemize}
\end{minipage} \hfill
\begin{minipage}[c]{0.2\textwidth}
\begin{center}
\includegraphics[width=\textwidth]{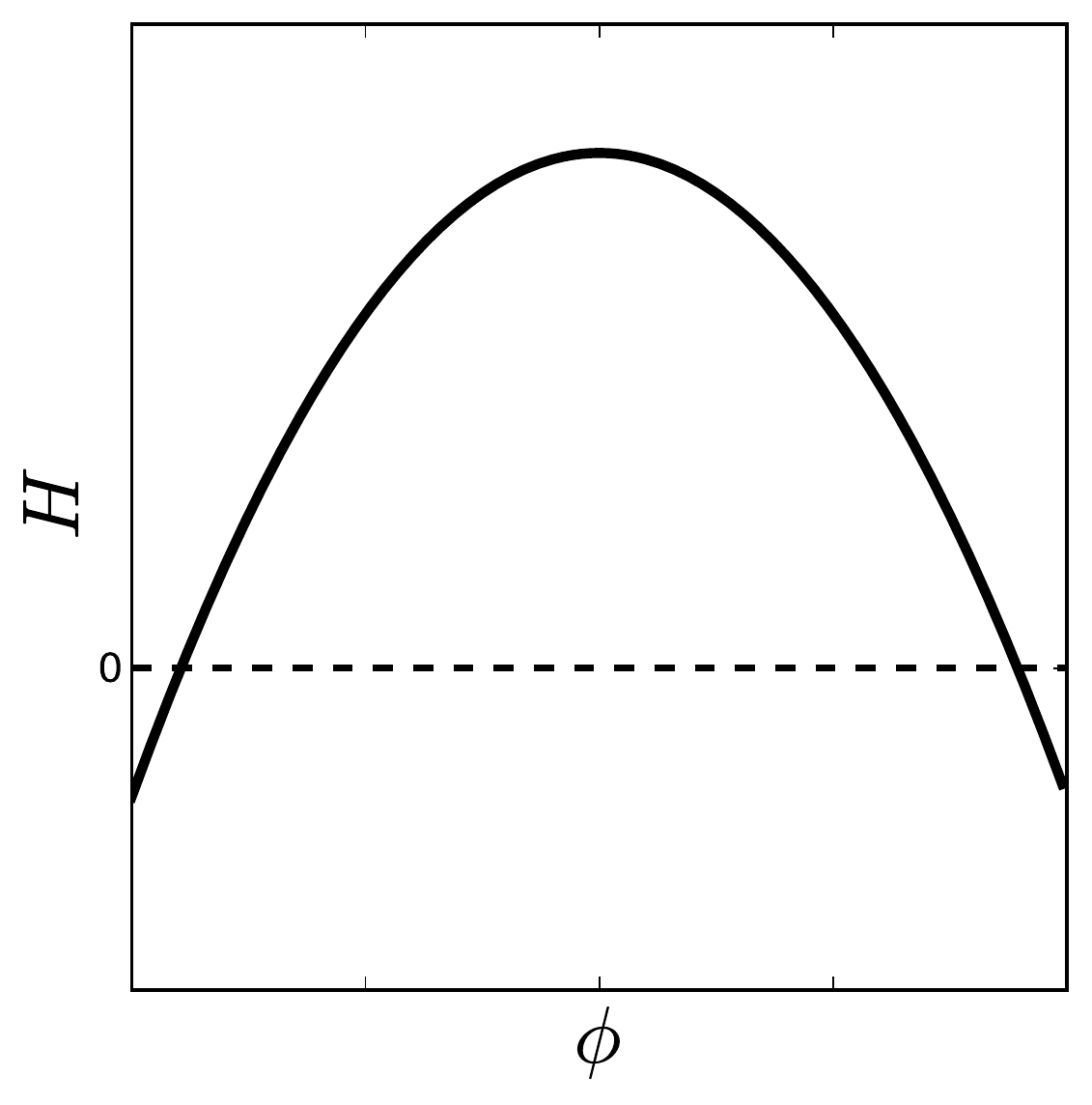}
\end{center}
\end{minipage}

\begin{minipage}{0.7\textwidth}
\begin{itemize}
\item
Secondly, if $\epsilon_1 < \epsilon_2 < 2 \epsilon_1$, then  both slow-roll parameters are vanishing in the far past and monotonically increasing during inflation. The corresponding Hubble function has a parabolic profile with a negative minimum (region II). In this case, $\gamma>0$ and $\xi=-1$.\\
\end{itemize}
\end{minipage} \hfill
\begin{minipage}[c]{0.2\textwidth}
\begin{center}
\includegraphics[width=\textwidth]{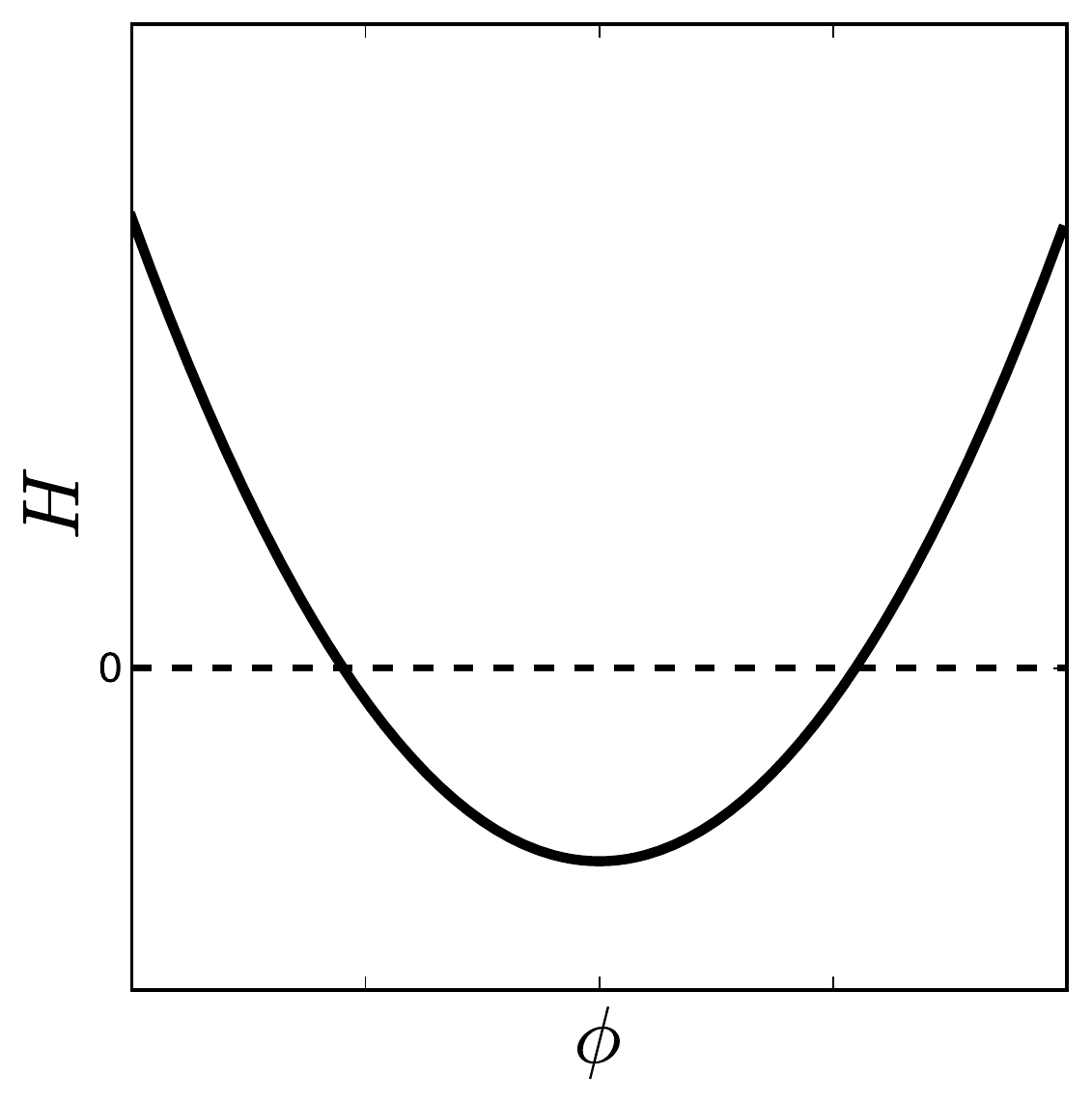}
\end{center}
\end{minipage}

\begin{minipage}{0.7\textwidth}
\begin{itemize}
\item
Thirdly and lastly, if $\epsilon_2 <\epsilon_1$, both slow-roll parameters are again vanishing in the far past. However, during inflation, $\epsilon_1$ reaches a maximum, and then decreases back to zero, while $\epsilon_2$ asymptotes to a negative value in the future. This stems from a Hubble function with a positive minimum (region III). In this case, $\gamma<0$, while $\xi=+1$ before $\epsilon_1$ crosses its maximum and $\xi=-1$ afterwards.\\
\end{itemize}
\end{minipage} \hfill
\begin{minipage}[c]{0.2\textwidth}
\begin{center}
\includegraphics[width=\textwidth]{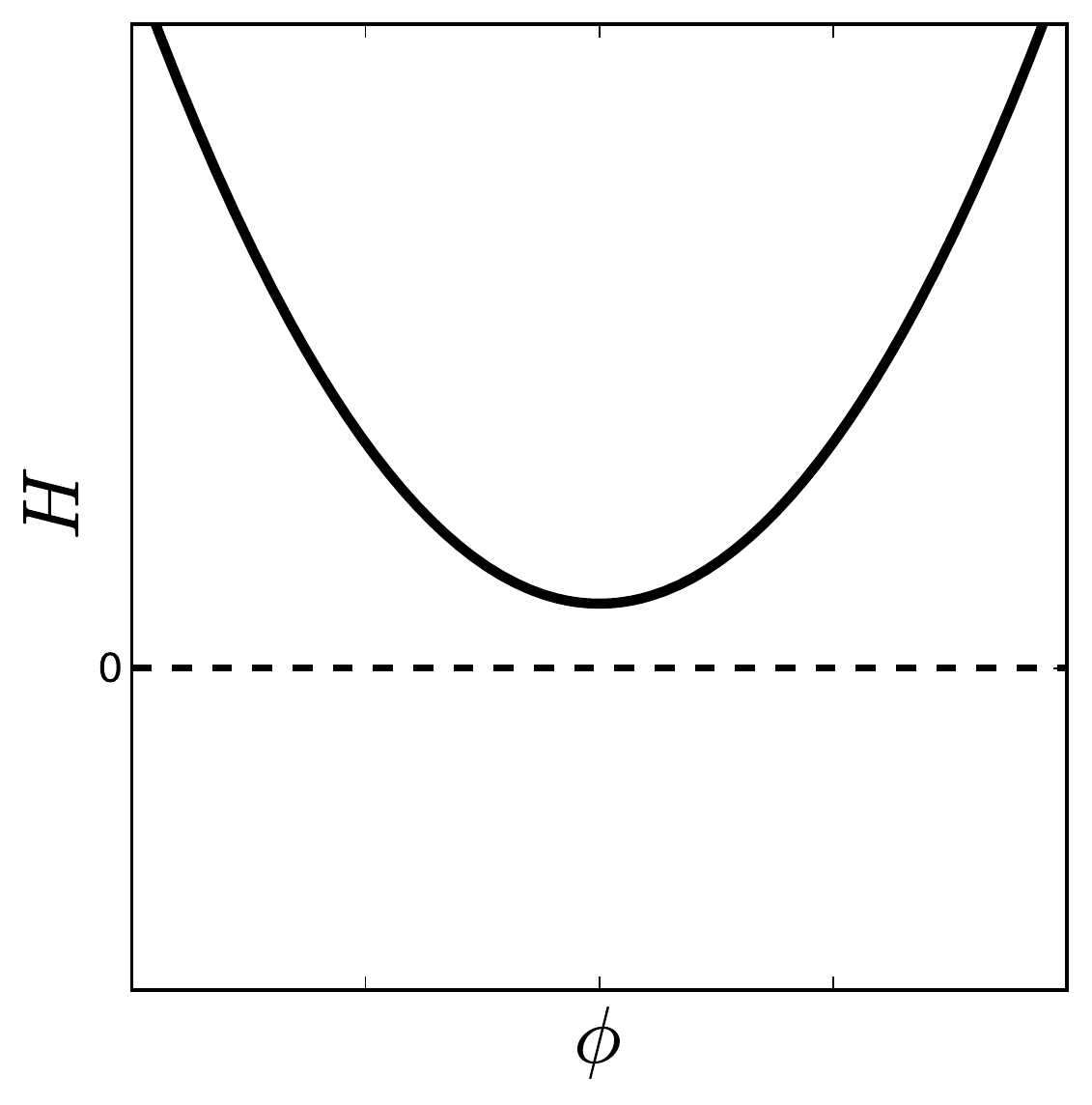}
\end{center}
\end{minipage}
These three regions are shaded with different colors in the left panel of \Fig{fig:Pade20}. One should note that thanks to the conservation of the sign of $\gamma$ defined in \Eq{eq:Taylor2:gamma:def}, a given inflationary trajectory never changes region. Amongst the third category, one can distinguish two cases. If the maximum value of $\epsilon_1$ is smaller than one, inflation never ends and reaches one of the fixed points $(\epsilon_1=0,\epsilon_2<0)$. If, on the other hand, the maximum value of $\epsilon_1$ is larger than one, and if one starts inflating with $\epsilon_2>0$, then inflation ends naturally when $\epsilon_1 = 1$. This happens when $\gamma<-4$.

Combining \Eqs{eq:epsNef:taylor}, (\ref{eq:eps2:feps}) and~(\ref{eq:nsr:srnlo}), the inflationary predictions of this class of models can be obtained and are displayed in the right panel of \Fig{fig:Pade20} for $40<\Delta N_*<60$. One should note that \Eqs{eq:nsr:srnlo} also make use of $\epsilon_3$, but $\epsilon_3$ is related to $\epsilon_1$ and $\epsilon_2$ thanks to \Eq{eq:eps3:Taylor}. For ``large-field'' scenarios (region II), $r$ is too large, and the model asymptotes the line $r= 16 (1-\nS)/3$ mentioned in section~\ref{sec:intro} and commented on in \Ref{Vennin:2014xta}, which separates regions II and III. For ``hilltop'' or ``small-field'' scenarios (region I), $r$ is small, but $\nS$ is generically too red. When interpolating between these two cases, there is a small range of models for which $r\sim 0.1$ and the spectral index $\nS$ has marginally the right value. However, one can check that this corresponds to very fine-tuned initial values of the flow parameters (or, equivalently, values of $\gamma$). Moreover, we will find in section~\ref{sec:numeric} that higher-order terms change this result significantly.
\subsection{\Pade Expansion}
\label{sec:analytics:exact:pad}
Let us then further illustrate our method by considering the case of a first order \Pade expansion,
\beq
   \frac{H}{H_0} = \frac{1 + a \phi}{1 + b \phi}  \,.
 \label{eq:H:Pade11}
\eeq
This case has not been considered in the literature before and provides a simple implementation of the idea of ``plateau inflation''. The first two slow-roll parameters can be obtained from \Eqs{eq:eps:H}, and one has
\beq
  \epsilon_1 = \frac{2 \left(a-b\right)^2}{\left(1+a \phi \right)^2 \left(1+b \phi\right)^2} \,, \quad \epsilon_2 =\frac{4 \left(a-b\right) \left(a+b+2 a b \phi\right)}{\left(1+a \phi\right)^2 \left(1+b \phi \right)^2}\, .
\eeq
Here, we follow exactly the same approach as the one used for the Taylor expansion in section~\ref{sec:analytics:exact:tay}. For instance, under shift transformations $\phi\rightarrow\phi+\delta\phi$, the functional form~(\ref{eq:H:Pade11}) is unchanged provided
\beq
  	a \to \frac{a}{1+a\delta\phi} \,, \quad 
	b \to \frac{b}{1+b\delta\phi} \,,
\eeq
where $H_0$ is also rescaled according to $H_0 \to H_0 (1+a\delta\phi)/(1+b\delta\phi)$. By moving to the gauge where the constant term in the numerator of \Eq{eq:H:Pade11} vanishes, \ie $\delta\phi=-1/a$, the $b$ coefficient becomes $b/(1-b/a)$, which is therefore gauge invariant. This implies that
\beq
\label{eq:def:gamma:Pade11}
	\gamma=\frac{16\sqrt{2}a b}{|a-b|}= \frac{\epsilon_2^2 - 4 \epsilon_1^2}{\epsilon_1^{3/2 }}
\eeq
is a constant of motion and can be used to label the different trajectories.

\begin{figure*}[t]
\begin{center}
\includegraphics[width=\widthdouble]{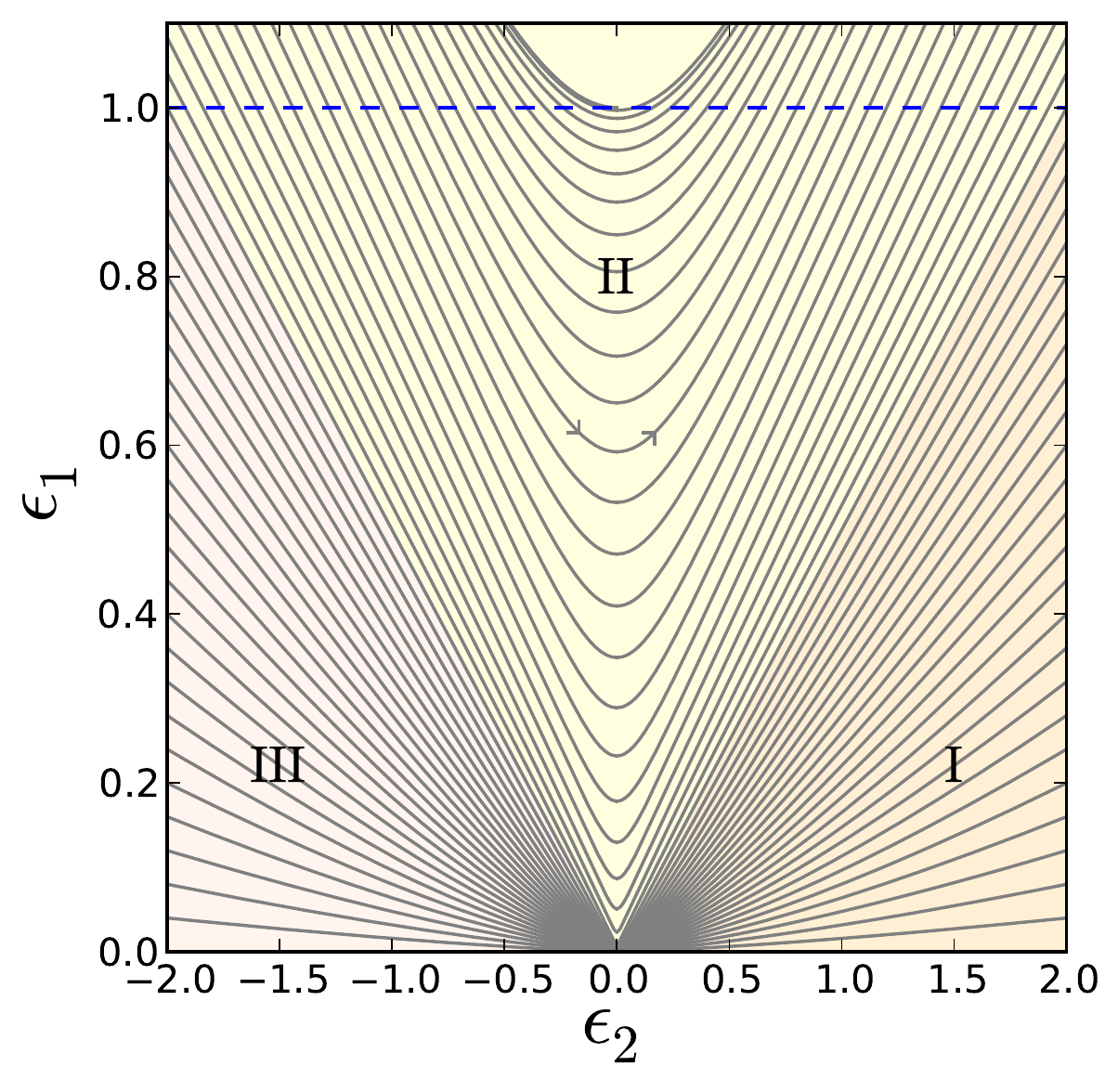}
\includegraphics[width=\widthdouble]{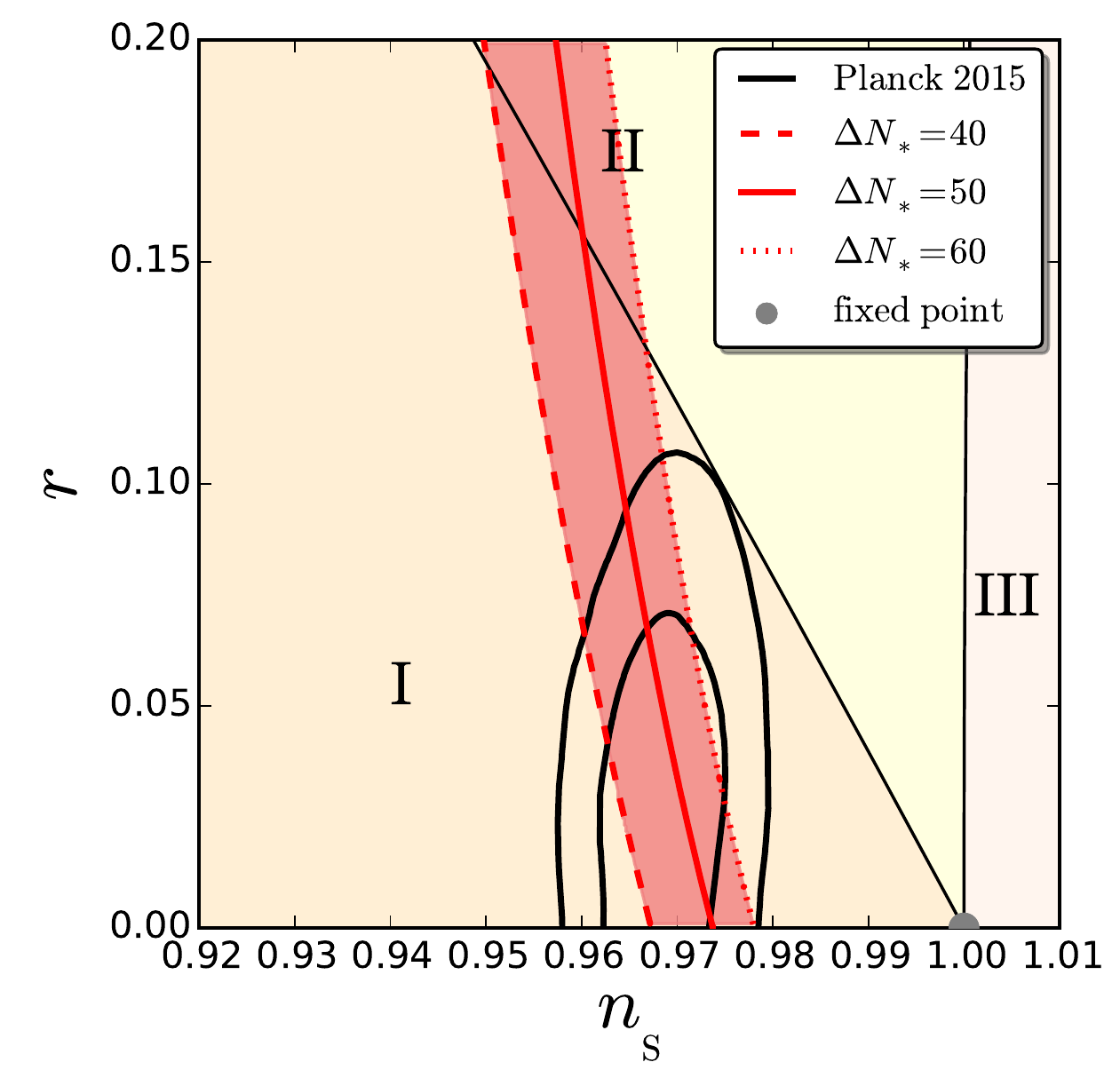}
\caption{\it First order \Pade expansion for the Hubble function. Left panel: flow lines of the system~(\ref{eq:depsdN:Pade11}) in the plane $(\epsilon_1,\epsilon_2)$. The arrows indicate in which direction inflation proceeds. The blue dashed line corresponds to $\epsilon_1=1$ where inflation stops. The three regions $\rm I$, $\rm II$ and $\rm III$ refer to the discussion in the main text. Right panel: Observational predictions in the $(\nS,r)$ plane, compared with the Planck 2015 $1\sigma$ and $2\sigma$ contours. The red lines stand for the values of $\nS$ and $r$ computed 40 \efolds before the end of inflation (dashed line), 50 \efolds (solid line) and 60 \efolds (dotted line). The grey dot at the bottom right stands for the fixed point $(\epsilon_1=0,\epsilon_2=0)$.}
\label{fig:Pade11}
\end{center}
\vspace{-0.5cm}
\end{figure*}

Let us recall that a given parametrization for the Hubble function can always be cast in a single dynamical system in the flow parameters space. For the present case, making use of the same procedure as before, \Eq{eq:H:Pade11} implies that $H^{\prime\prime\prime}=3(H^{\prime\prime})^2/(2H^\prime)$, and \Eqs{eq:eps:H} give rise to
\beq
\label{eq:eps3:Pade}
   \epsilon_3 = \frac{\epsilon_1^2}{\epsilon_2} + \frac34\epsilon_2 \,.
\eeq
Again, this truncates the dynamical system to a closed set of differential equations for $(\epsilon_1, \epsilon_2)$, given by
\beq
\label{eq:depsdN:Pade11}
   \frac{\dd \epsilon_1}{\dd N} = \epsilon_1 \epsilon_2 \,, \qquad
   \frac{\dd \epsilon_2}{\dd N} = \epsilon_1^2  + \frac34 \epsilon_2^2 \,. 
\eeq
In particular, one can check that the combination $\gamma$ defined in \Eq{eq:def:gamma:Pade11} is left invariant. The equation for $\epsilon_1$ is the same as in \Eq{eq:Taylor2:floweqs}, since it just defines $\epsilon_2$. However, the equation for $\epsilon_2$ is different. In general indeed, only the flow equation for the last flow parameter encodes the physical information about the model, and propagates back to yield a specific dynamics for all flow parameters. The integrated flow lines of the above system are displayed in the left panel of \Fig{fig:Pade11}. 

Let us now see how this system can be integrated analytically. By inverting \Eq{eq:def:gamma:Pade11}, one can express $\epsilon_2$ as a function of $\epsilon_1$,
\beq 
\label{Pade-constant}
   \epsilon_2 = \xi \sqrt{ 4 \epsilon_1^2 + \gamma \epsilon_1^{3/2}} \,,
\eeq
where $\xi=\pm 1 =\mathrm{sign}(\epsilon_2)$ changes when $\epsilon_1$ crosses its minimum value. As before, inserting \Eq{Pade-constant} into \Eqs{eq:depsdN:Pade11} yields a first order differential equation for $\epsilon_1(N)$ that can be solved, and one obtains
\beq
\label{eq:Pade11:nef:2}
 \Delta N_*=N(\epsilon_{1,\mathrm{end}})-N(\epsilon_{1*}) \,, \qquad N \left(\epsilon_1\right) = \xi \frac{4\left(8\sqrt{\epsilon_1}- \gamma \right)\sqrt{4{\epsilon_1}+ \gamma }}{3\epsilon_{1}^{3/4} \gamma^2}\, ,
\eeq
where again $\epsilon_{1,\mathrm{end}}=1$. Contrary to the result obtained in section~\ref{sec:analytics:exact:tay}, this expression cannot be inverted analytically.

Let us now discuss the structure of the phase space diagram ploted in the left panel of \Fig{fig:Pade11}. According to the type of Hubble function, three possibilities must again be distinguished:\\

\begin{minipage}{0.7\textwidth}
\begin{itemize}
\item
Firstly, if $\epsilon_2 > 2 \epsilon_1$, both $\epsilon_1$ and $\epsilon_2$ increase as inflation proceeds, from the (repulsive) fixed point $\epsilon_1 = \epsilon_2 =0$ reached in the  infinite past. Inflation ends naturally when $\epsilon_1=1$. This implements the idea of ``plateau inflation'' where the Hubble function is concave with a non vanishing plateau where inflation proceeds (region I). In this case, $\gamma>0$ and $\xi=+1$.\\
\end{itemize}
\end{minipage} \hfill
\begin{minipage}[c]{0.2\textwidth}
\begin{center}
\includegraphics[width=\textwidth]{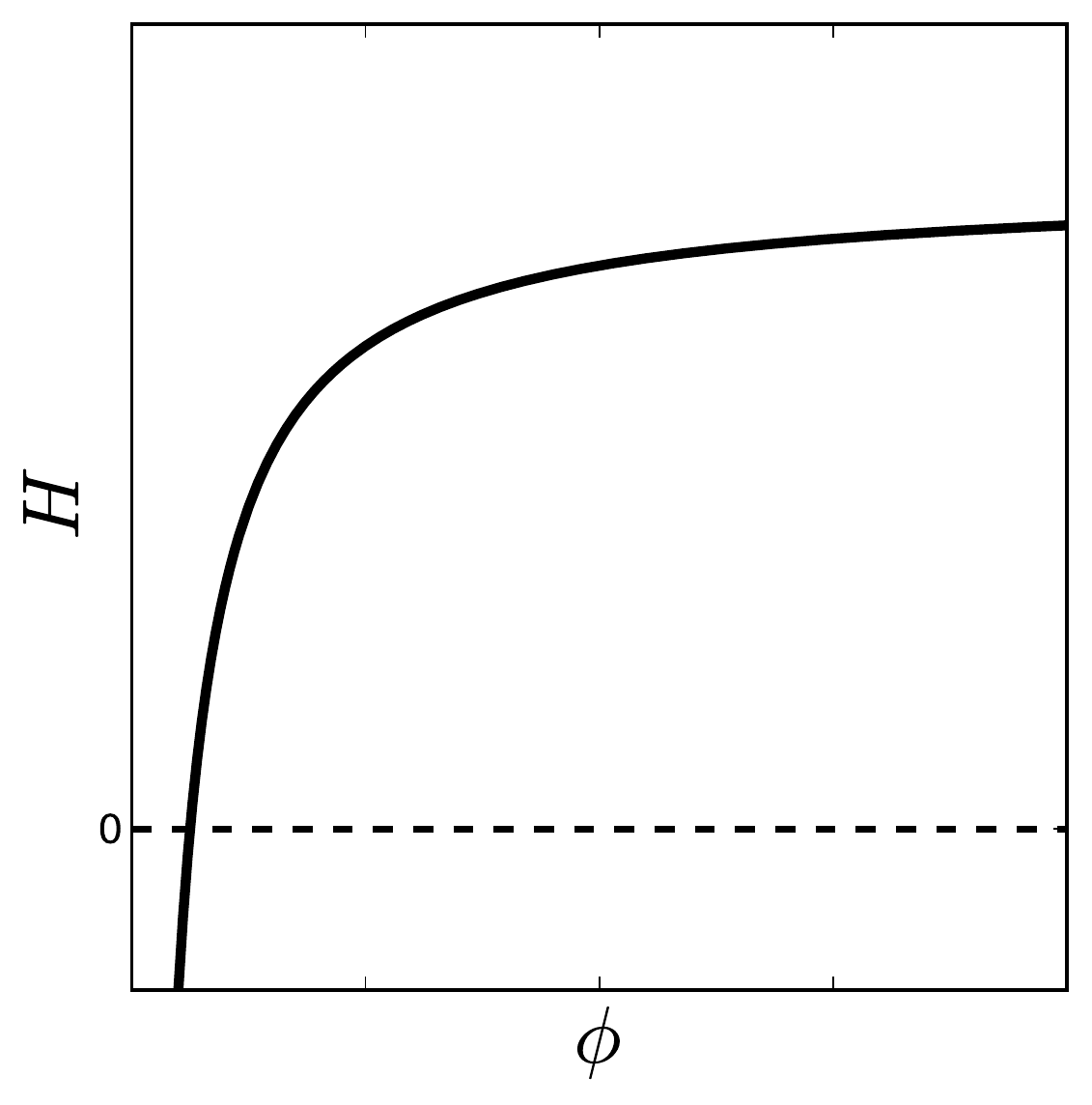}
\end{center}
\end{minipage}

\begin{minipage}{0.7\textwidth}
\begin{itemize}
\item
Secondly, if $-2\epsilon_1 < \epsilon_2 < 2 \epsilon_1$, $\epsilon_1=1$ is reached both in the past and in the future. In between, a finite period of inflation takes place where $\epsilon_2$ increases and $\epsilon_1$ goes through a minimum. The corresponding Hubble function is convex and vanishes before the plateau is reached (region II). In this case, $\gamma<0$ while $\xi=-1$ before $\epsilon_1$ reaches its minimum, and $\xi=-1$ afterwards.\\
\end{itemize}
\end{minipage} \hfill
\begin{minipage}[c]{0.2\textwidth}
\begin{center}
\includegraphics[width=\textwidth]{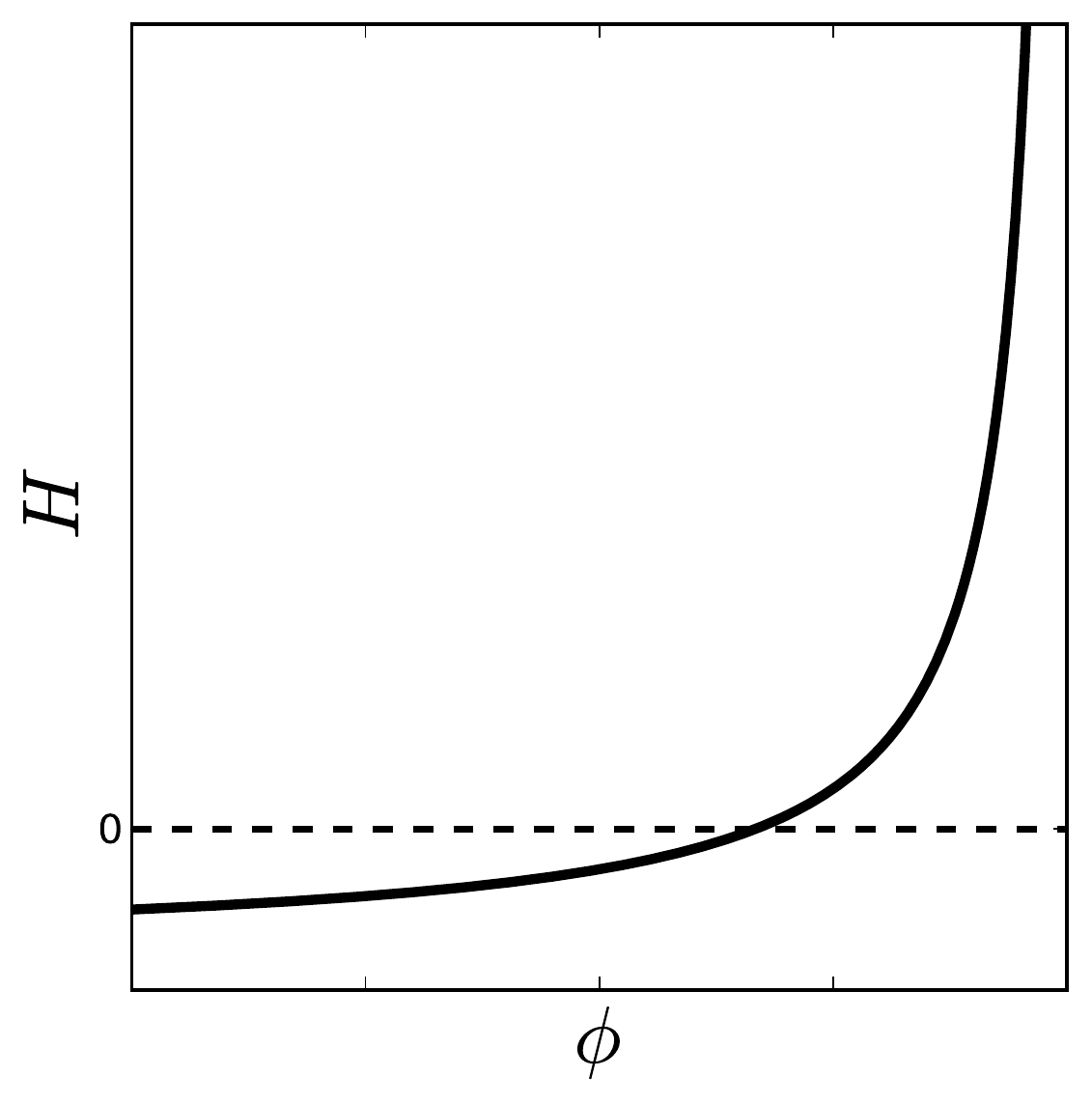}
\end{center}
\end{minipage}

\begin{minipage}{0.7\textwidth}
\begin{itemize}
\item
Thirdly and lastly, if $\epsilon_2 < - 2 \epsilon_1$, $\epsilon_1$ decreases as inflation proceeds, while $\epsilon_2$ increases. The (attractive) fixed point $\epsilon_1 = \epsilon_2=0$ is reached in the asymptotic future. This corresponds to a convex Hubble function for which the plateau is positive (region III). In this case, $\gamma>0$ and $\xi=-1$.\\
\end{itemize}
\end{minipage} \hfill
\begin{minipage}[c]{0.2\textwidth}
\begin{center}
\includegraphics[width=\textwidth]{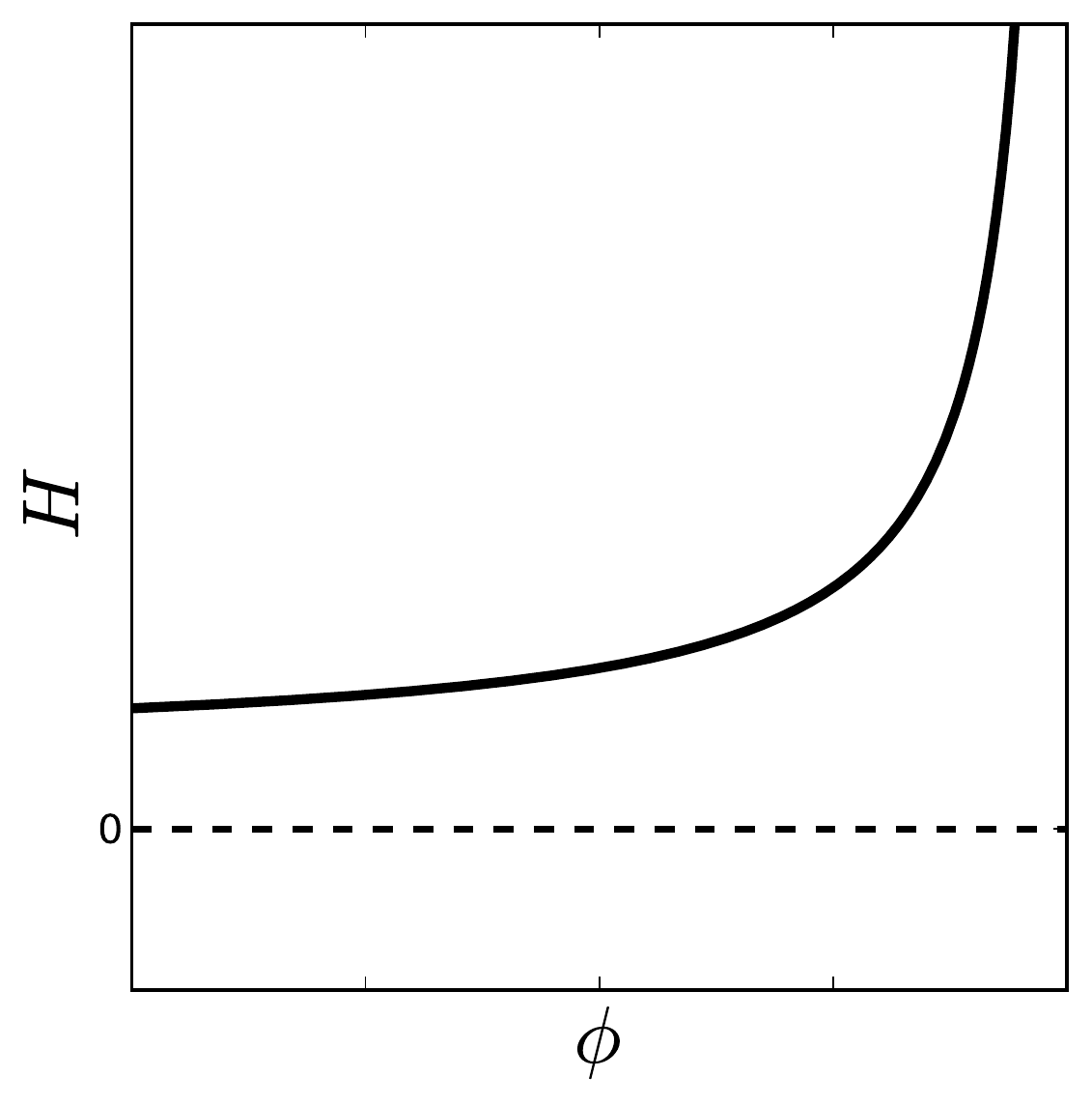}
\end{center}
\end{minipage}
In particular, making use of \Eq{eq:Pade11:nef:2}, one can check that an infinite number of \efolds can be realized in cases I and III. However, in case II, only a finite number of \efolds can be obtained. Parametrizing a given trajectory within region II by $\epsilon_{2,\uend}=\sqrt{\gamma+4}$, the value of the second flow parameter at the end of inflation, this number is given by
\beq
N_{\mathrm{max}} = \frac{8}{3}\epsilon_{2,\mathrm{end}}\frac{12-\epsilon_{2,\mathrm{end}}^2}{\left(\epsilon_{2,\mathrm{end}}^2-4\right)^2} \,.
\eeq
As expected, this number vanishes when $\epsilon_{2,\mathrm{end}}$ approaches $0$ and diverges when $\epsilon_{2,\mathrm{end}}$ approaches $2$.

Combining \Eqs{eq:Pade11:nef:2}, (\ref{Pade-constant}) and~(\ref{eq:nsr:srnlo}), the inflationary predictions of this class of models can be obtained and are displayed in the right panel of \Fig{fig:Pade11} for $40<\Delta N_*<60$. Again, in \Eqs{eq:nsr:srnlo}, $\epsilon_3$ is related to $\epsilon_1$ and $\epsilon_2$ thanks to \Eq{eq:eps3:Pade}. When inflation proceeds in region I, in the limit $\epsilon_{2\uend}\gg 1$, one recovers the ``typical'' predictions of plateau inflation where $r$ is small and $\nS$ is in good agreement with the observational constraints. In this limit, \Eq{Pade-constant} gives rise to $\epsilon_2 \simeq \sqrt{\gamma} \epsilon_1^{3/4}$, and one has
\beq
  \epsilon_{1*} \simeq\left( \frac{4}{3 \sqrt{\gamma} \Delta N_*} \right)^{4/3} \,, \qquad
 \epsilon_{2*} \simeq \frac{4}{3 \Delta N_*} \,.
\eeq
This translates into $\nS\simeq 1-4/(3\Delta N_*)$ and $r \sim \Delta N_*^{-4/3} \ll 1$, which is what one would expect from a  plateau inflation model with a $1/\phi$ fall-off \cite{Mukhanov:2013tua, Roest:2013fha}.

Finally, let us note that this regime is interesting because $\epsilon_{2\uend}\gg 1$ means that the last stage of the inflationary phase is realized far away from slow roll (let us recall that, here, the inflationary dynamics is solved \emph{without} resorting to the slow-roll approximation). However, the number of \efolds realized between the time when $\epsilon_2=1$ and the end if inflation when $\epsilon_1=1$ can be calculated thanks to \Eq{eq:Pade11:nef:2}, and in the limit where $\epsilon_{2\uend}\gg 1$, one obtains $4/3$ $e$-folds. This is why, $\Delta N_*$ \efolds before the end of inflation, slow roll is well valid and the system gives rise to predictions that are in good agreement with observations. 

\subsection{Comparison}

\begin{figure*}[t]
\begin{center}
\includegraphics[width=8.5cm]{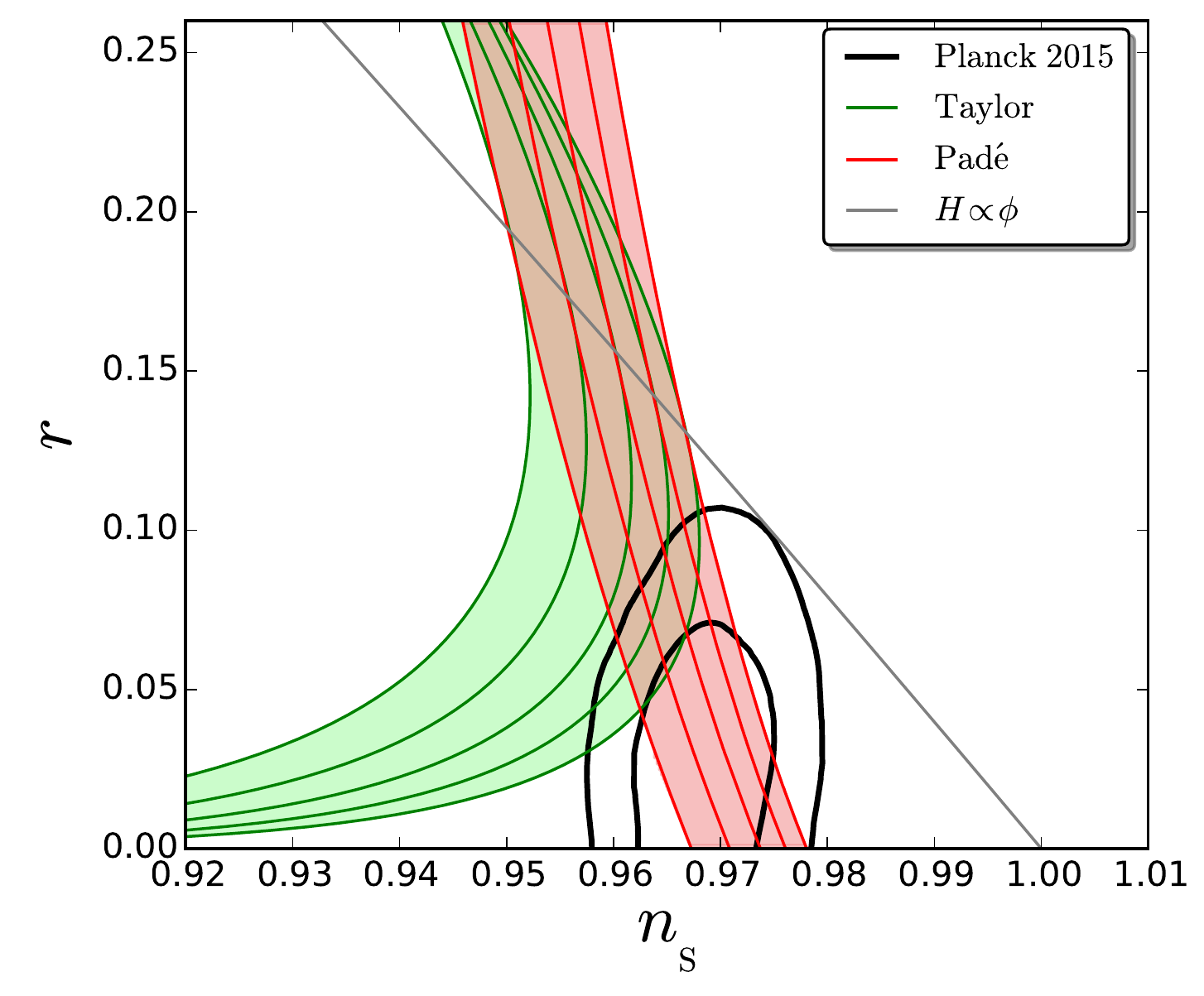}
\caption{\it Compared predictions of the Taylor model~(\ref{H-Taylor}) (green) and the \Pade model~(\ref{eq:H:Pade11}) (red), for $40<\Delta N_*<60$, in the $(\nS,r)$ plane. The black lines are the $1\sigma$ and $2\sigma$ contours of Planck 2015. The grey solid line stands for the model $H/H_0 = \phi$, which is a special case of both parametrizations and for which $\epsilon_3=\epsilon_2=2\epsilon_1$.}
\label{fig:ComparedPred}
\end{center}
\vspace{-0.5cm}
\end{figure*}

In order to summarize the analysis of the two toy models discussed in the present section, in \Fig{fig:ComparedPred}, we have superimposed their predictions in the $(\nS,r)$ plane, for a few fixed values of $\Delta N_*\in [40,60]$. One can see that Taylor and \Pade lines are tangential, along the $\epsilon_3=\epsilon_2=2\epsilon_1$ line, which is associated to the model $H/H_0 = \phi$. This should not come as a surprise for the following reason. After a suitable gauge transformation, \Eq{H-Taylor} can be cast in the form 
 \begin{align} 
\frac{H}{H_0} =\phi+\sqrt{\frac{\gamma}{2}}\frac{\phi^2}{4} \,,
 \end{align}
where $\gamma$ has been defined in \Eq{eq:Taylor2:gamma:def}. Similarly, \Eq{eq:H:Pade11} can be cast in the form 
 \begin{align}
\frac{H}{H_0} =\frac{\phi}{1+\frac{\gamma}{16\sqrt{2}}\phi} \,,
\end{align} 
where $\gamma$ has been defined in \Eq{eq:def:gamma:Pade11}.  As a consequence,  a linear Hubble is a special case of both parametrizations, corresponding to $\gamma=0$. However, obviously, the way $\gamma$ modifies this linear $H(\phi)$ function is different for both parametrizations. 
\section{Numerical Integration of the Hubble Flow}
\label{sec:numeric}
The above results indicate that inflationary dynamics is better parametrized by \Pade expansions of the Hubble function rather than Taylor expansions. However, one might worry that this statement relies on the low truncation order we have worked with. This is why in this section, we generalize our approach by including higher order terms numerically. This also allows us to investigate a crucial aspect of flow parametrizations, namely the dependence (or lack thereof) of the results on the prior choice for the parameters of the expansion. Notice also that the analytical method developed in section~\ref{sec:analytics} can in principle be used to deal with arbitrarily large order expansions, however, one would not gain much by displaying the corresponding cumbersome formulas. This is why, here, we directly compute the predictions of the models we study, which consist in Hubble functions of the form
\beq
\label{Pade}
H\left(\phi\right) = H\left[m,n\right]\left(\phi\right) \equiv \displaystyle\frac{ \sum_{k=0}^m c_k \phi^k}{\sum_{l=0}^n d_l \phi^l} \,.
\eeq
In practice, we consider orders $[M,0]$, which correspond to Taylor expansions of the Hubble function, and orders $[M,M]$, which correspond to Hubble functions that asymptotes to a non-vanishing plateau at large-field values. We study different values of $M$ in order to test the robustness of the predictions under increasing the order of truncation and we report the results below. 

When the Hubble function is described by a Taylor series, we start our exploration at $\phi_\mathrm{ex}=0$. This is because, in principle, the Taylor expansion is defined around $\phi=0$, and might not always converge far from this value. A \Pade approximant is a simultaneous expansion around $\phi=0$ and $\phi=\infty$, which allows us to either start the exploration of the \Pade Hubble function at $\phi_\mathrm{ex}=0$ or $\phi_\mathrm{ex}=\infty$. This last case corresponds to plateau inflation and we expect it to be in best agreement with the data. A first flow algorithm was designed in \Ref{Kinney:2002qn}, and here we use the modification proposed in \Ref{Ramirez:2005cy} that allows us to include generic Hubble functions. It proceeds according to the following steps:
\begin{itemize}
\item[(i)] Draw the parameters $c_k$ and $d_k$ appearing in \Eq{Pade} according to some prior distribution (see below).
\item[(ii)] The Hubble function being specified, calculate $\epsilon_1(\phi)$ through \Eq{eq:eps:H}.
\item[(iii)] If $\epsilon_1(\phi_\mathrm{ex})>1$, restart from (i). Otherwise go to (iv).
\item[(iv)] Calculate the set of values $\lbrace \phi_0\rbrace $ where $\epsilon_1=0$ and $\lbrace \phi_1\rbrace $ where $\epsilon_1=1$.
\item[(v)] If $H^\prime(\phi_\mathrm{ex})>0$, inflation proceeds at decreasing values of $\phi$. Amongst the elements of $\lbrace \phi_0\rbrace$ and $\lbrace\phi_1\rbrace$, find which value is the closest to $\phi_\mathrm{ex}$ while being smaller than $\phi_\mathrm{ex}$. If this value belongs to $\lbrace \phi_0\rbrace$, a fixed point is reached and we chose not to consider such trajectories in the present analysis since the results then depend on the value of $\phi$ at which inflation terminates, which has to be determined by other physical processes than slow-roll violation. If, on the other hand, it belongs to $\lbrace \phi_1 \rbrace$, identify $\phi_\uend$ with this value. Apply a similar procedure if $H^\prime(\phi_\mathrm{ex})<0$. 
\item[(vi)] Calculate the value $\phi_*$ of $\phi$, $\Delta N_*=50$ \efolds prior to $\epsilon_1=1$, by integrating
\beq
	\Delta N_*=\sign(H^\prime) \int_{\phi_*}^{\phi_\uend}\frac{\dd \phi}{\sqrt{2 \epsilon_1}}
\eeq
and inverting the result. If a sufficient number of \efolds cannot be obtained, go back to (i).
\item[(vii)] Calculate the slow-roll parameters $\epsilon_{1*}$, $\epsilon_{2*}$ and $\epsilon_{3*}$ making use of \Eqs{eq:eps:H}, and the values for $\nS$ and $r$ thanks to \Eqs{eq:nsr:srnlo}.
\end{itemize}
In practice, we iterate this procedure until we obtain $10^6$ successful realizations.
Note that the results presented below have also been derived with different values of $\Delta N_*$, and that we have checked that our conclusions remain unchanged.

In step (i), the coefficients of the expansion are drawn according to some priors that we now specify. We studied two classes of priors. The first one consists in drawing all coefficients $c_k$ and $d_k$ from a flat distribution between $[-p/q^k,p/q^k]$, where $p$ and $q$ are two fixed numbers. In the following, it is referred to as the ``power-law'' priors. In the second class of priors, $c_k$ and $d_k$ are drawn from flat distributions in the interval $[-p f_k, p f_k]$, where $p$ is a constant and the set $\lbrace f_k \rbrace$ is defined such that if $c_k=(-1)^k d_k=f_k$ for all $k$, all $M$ derivatives of the Hubble function at $\phi=0$ are $1$. This gives rise to
\beq
\label{eq:Pade:par}
	{\rm Taylor:~~} f_k=\frac{1}{k!} \,, \qquad {\rm Pade:~~} f_k=\displaystyle\frac{\binom{M}{k}}{\binom{2M}{k}k!}.
\eeq
%This prescription ensures that the randomly drawn derivatives are of the same order. 
In what follows, this prescription is referred to as the ``binomial priors''.

\begin{figure*}[t]
\includegraphics[width=0.5\columnwidth]{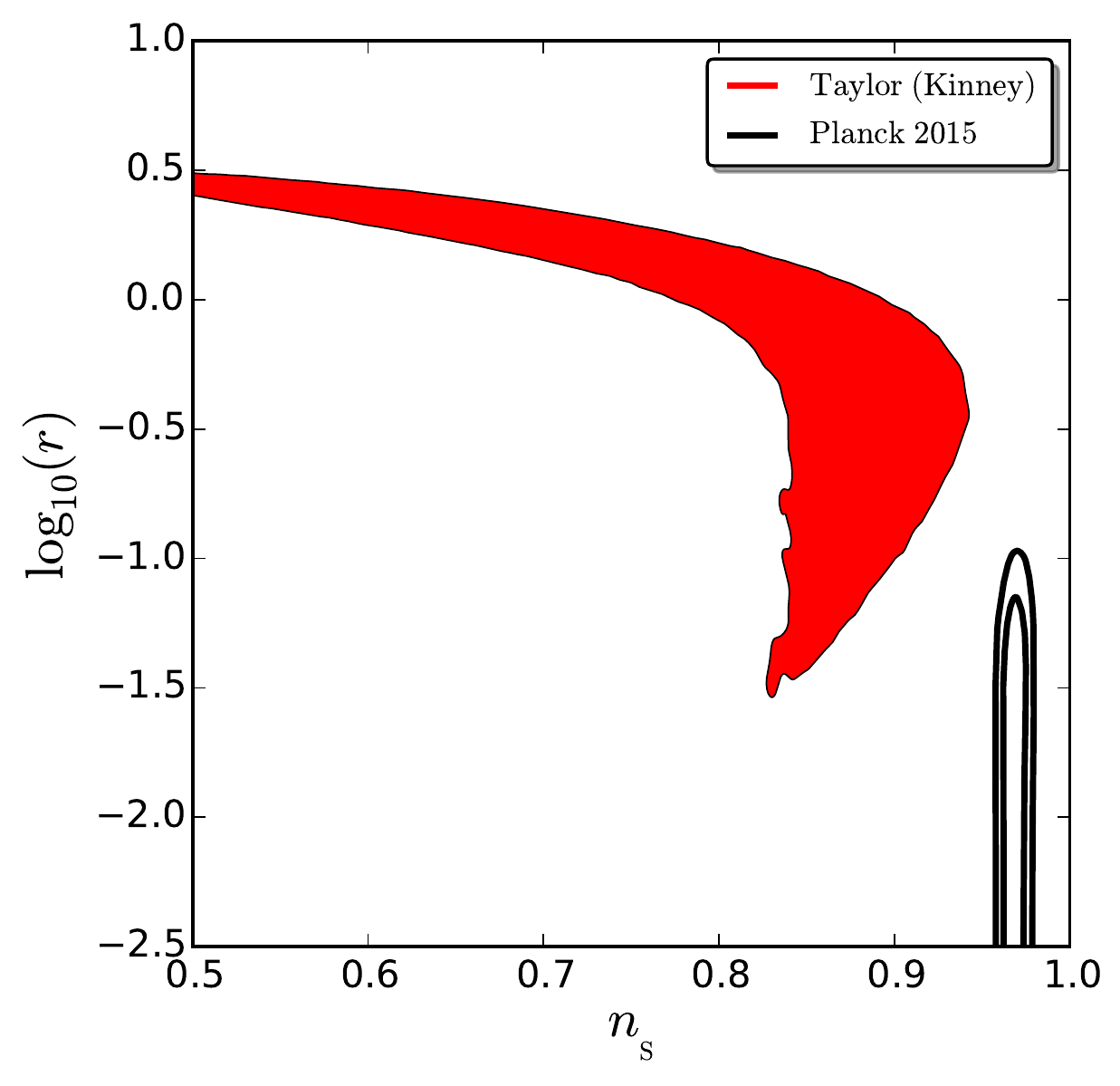}
\includegraphics[width=0.5\columnwidth]{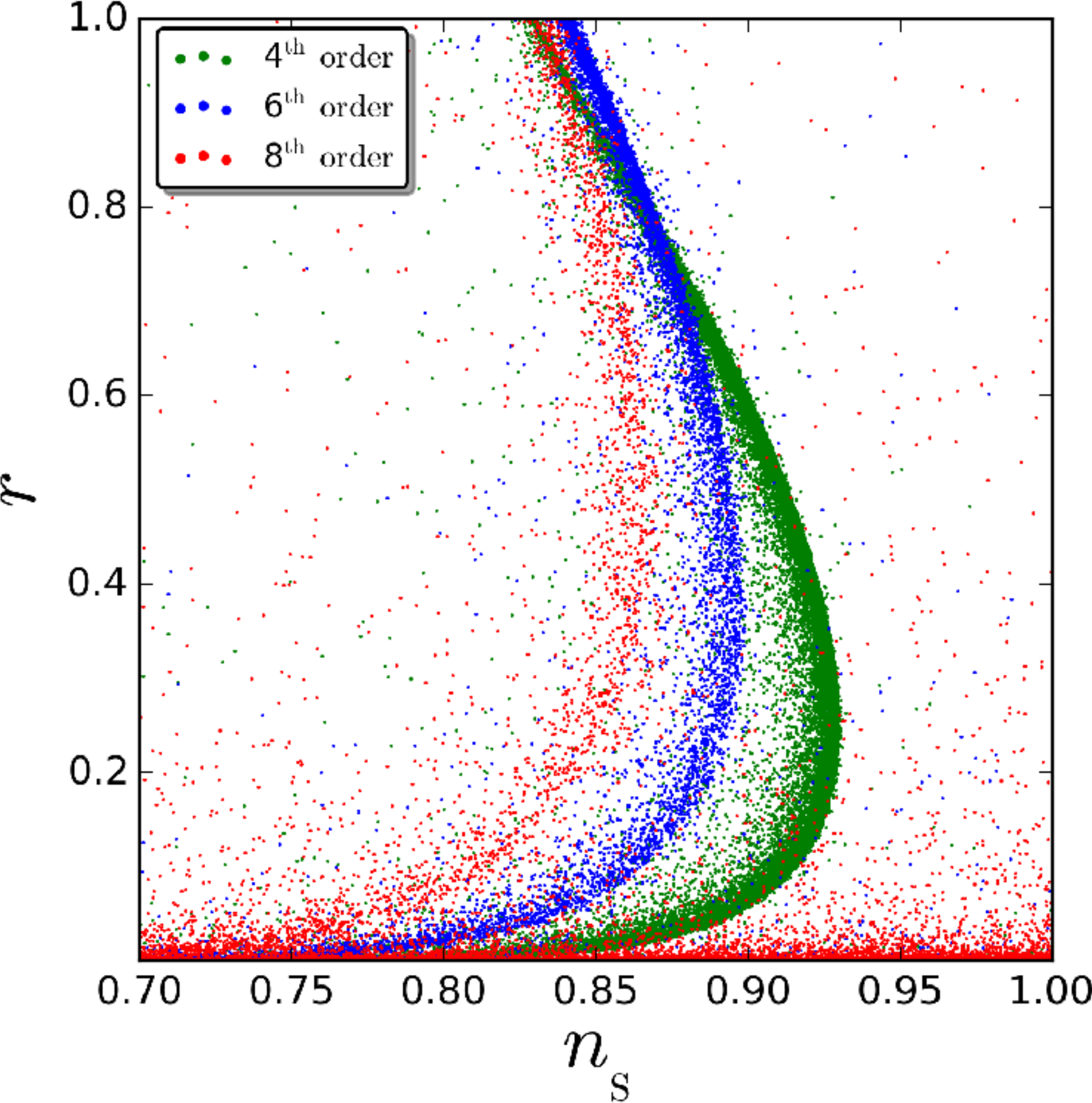}
  \caption{\it Left panel: one sigma contour of the predictions generated by a Taylor expansion of the Hubble function, with priors on the coefficients of the expansion matching the procedure originally proposed in \Ref{Kinney:2002qn}. Far from $\nS=1$ and $r=0$, the slow-roll formulas~(\ref{eq:nsr:srnlo}) are not valid and we display the contour for indicative purpose only. The Planck $1$ and $2\sigma$ contours are shown for comparison. Right panel: prediction points for a Taylor expansion, truncated at different orders $M$, when the coefficients are drawn from the binomial prior distributions.}
  \label{fig:numerics:Taylor}
\vspace{-0.5cm}
\end{figure*}

The original procedure proposed by Kinney \cite{Kinney:2002qn} relies on drawing $ c_k \in(\sqrt{2}/((2q)^k(k+1)!))[-p,p]/[0,1]^{(k-1)/2}$, with $d_0=1$ and $d_k=0$ for $k\geq1$, where $[-p,p]/[0,1]^{(k-1)/2}$ stands for the ratio of two numbers, one drawn in the interval $[-p,p]$ and the other one drawn in $[0,1]$ and taken to the $(k-1)/2$ power. This corresponds to taking power-law priors on the values of the $\lambda$ flow parameters, defined in \Eq{eq:lambda:def}, at $\phi_\mathrm{ex}=0$. The $1\sigma$ contour obtained with this prescription is shown in the left panel of \Fig{fig:numerics:Taylor}. It is widely spread, and since the $2\sigma$ contour is even more spread and difficult to resolve statistically, we do not display it. From the Planck contours, one can see that the vast majority of trajectories generated by this procedure are observationally excluded. This can be further quantified by calculating the percentage of the points lying outside the Planck $2\sigma$ contour. One obtains $0.2$\%, a small fraction indeed. We should stress that these results correspond to the standard ``horizon flow'' procedure as commonly used in the literature. They again motivate our search for alternative parametrizations.

Unfortunately, other natural prior choices for the Taylor expansion coefficients, such as the power-law or the binomial priors introduced above, fail to converge as $M$ increases. In other words, the obtained results have a large dependence on the order of truncation, an undesirable property. More precisely, in the power-law case, since the radius of convergence of the Taylor series is $\phi=q$, the model converges only when $q$ is sufficiently large. For such priors, the higher order terms are strongly suppressed and, in practice, the simple results of the second order truncation obtained in section~\ref{sec:analytics:exact:tay} are recovered, and the higher order information is quickly lost. In the binomial case, the results corresponding to different orders of truncation are displayed in the right panel of \Fig{fig:numerics:Taylor}, where it is obvious that the model does not converge. Therefore, it seems difficult to design alternative parametrizations relying on a Taylor expansion.

\begin{figure*}[t]
\includegraphics[width=0.5\columnwidth]{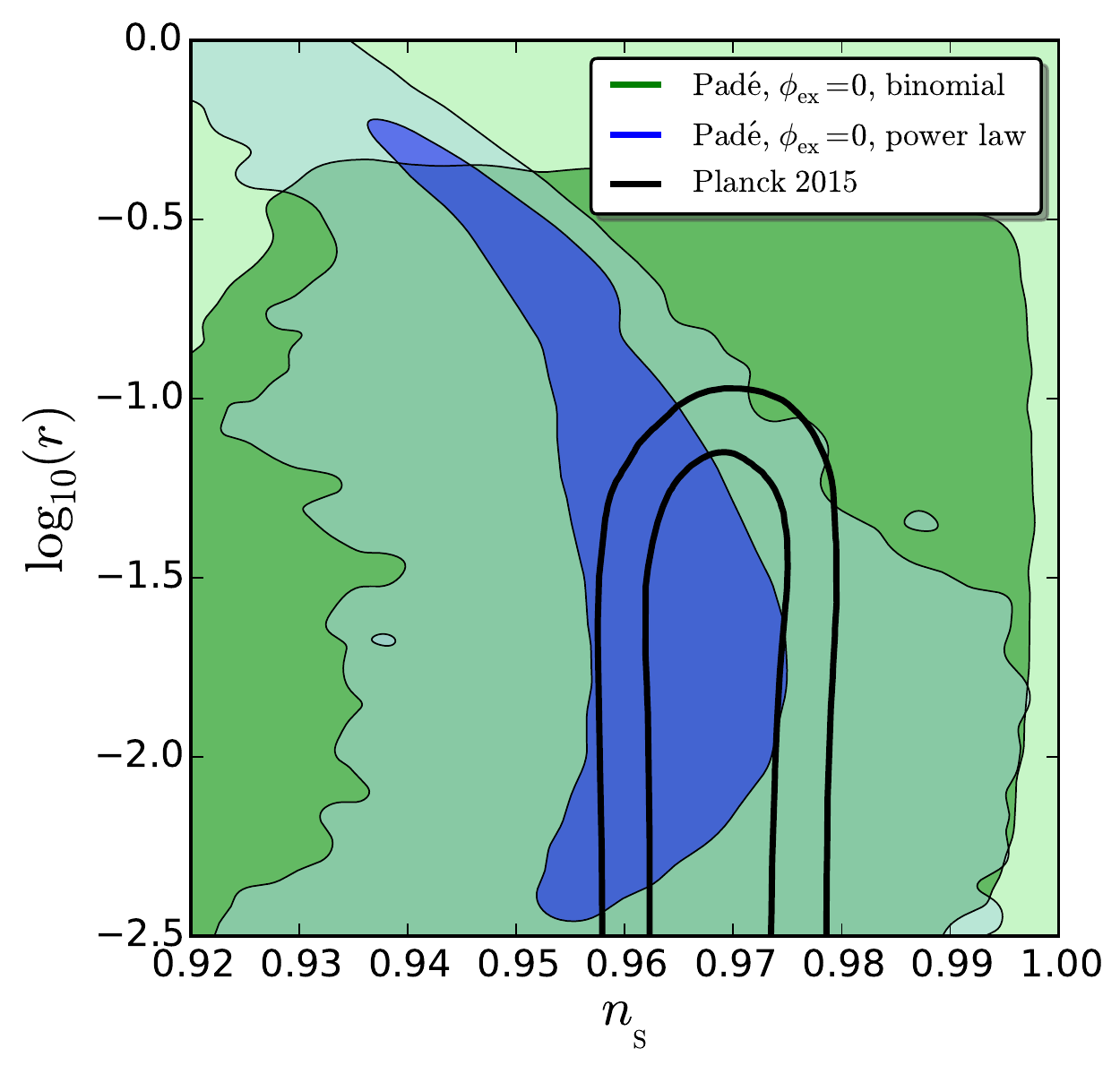}
\includegraphics[width=0.5\columnwidth]{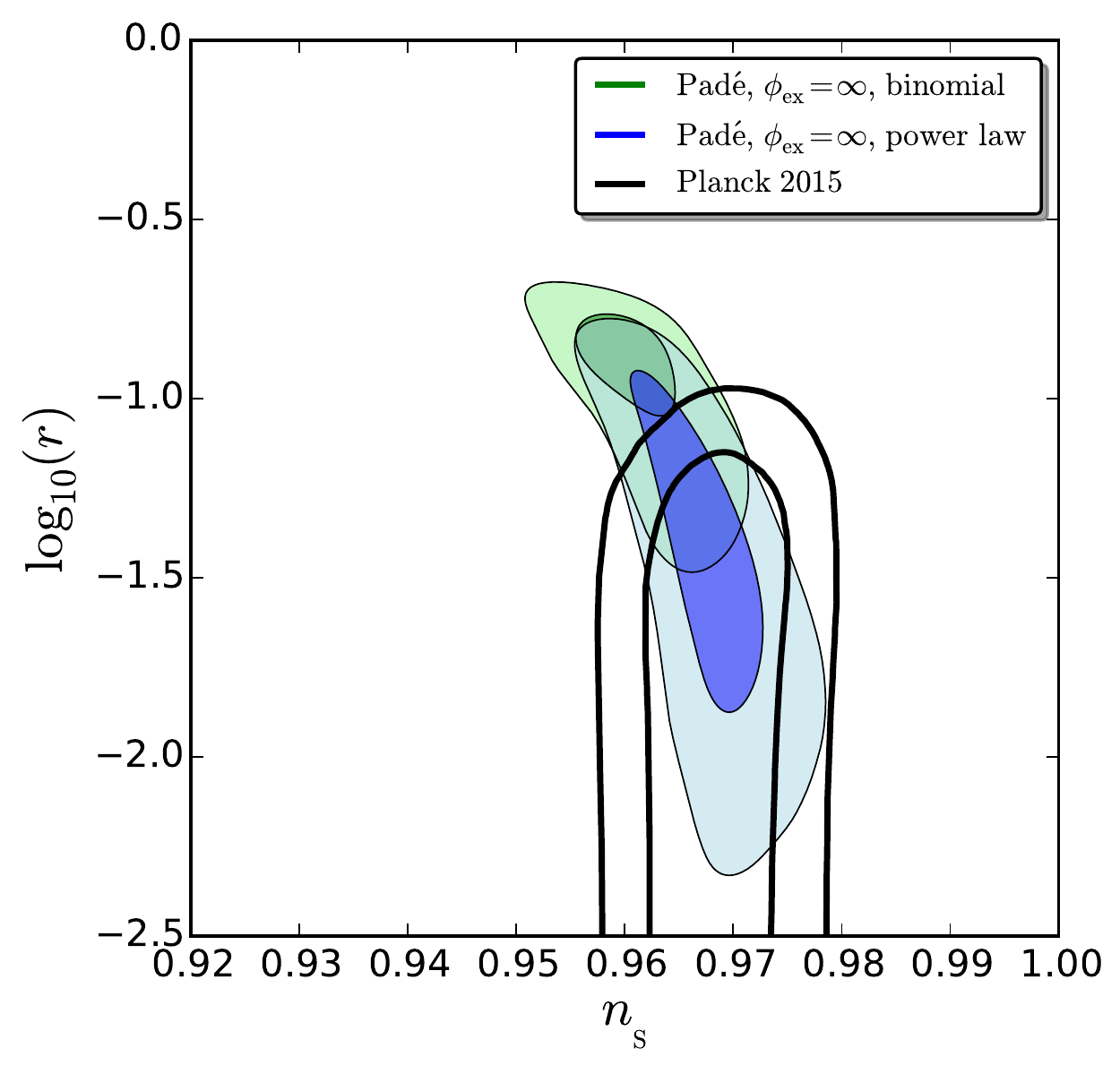}
  \caption{\it $1$ and $2\sigma$ contours for the two prior choices (power law and binomial), for a \Pade expansion of the Hubble function. The left panel corresponds to $\phi_{\mathrm{ex}}=0$ and the right panel to $\phi_\mathrm{ex}=\infty$.}
  \label{fig:Padeprior}
  \vspace{-0.5cm}
\end{figure*}

Let us now consider \Pade approximants. First of all, we have checked that with the two classes of priors proposed above, the results always converge (indeed, it is well known that \Pade approximants have better convergence properties than Taylor expansions). In practice, we find that $M=6$ is enough to reproduce all higher order results with a very good accuracy. The next question is how much the results depend on the class of priors. In \Fig{fig:Padeprior}, we have displayed the $1$ and $2\sigma$ contours obtained with the power-law and binomial priors, in the case where $\phi_\mathrm{ex}=0$ (left panel) and $\phi_\mathrm{ex}=\infty$ (right panel). 

When $\phi_\mathrm{ex}=0$, the binomial prior gives rise to wide spread contours (the $2\sigma$ contour entirely lies outside the plot frame). They are consistent with the power-law results, but scan larger sets of inflationary trajectories. For this reason, the percentage of points inside the Planck $2\sigma$ contour is smaller, $5$\% for the binomial distribution and $38$\% with the power-law prior.

When $\phi_\mathrm{ex}=\infty$, we find that both priors give rise to rather narrow contours, in agreement with each other at the $1\sigma$ level. In this case, large fractions of points lie inside the Planck $2\sigma$ contour: $18$\%  for the binomial prior choice and $90$\% when the power-law prior is used. These models correspond indeed to one's intuitive representation of ``plateau inflation''. Interestingly also, a lower bound on $r$ is found, which means that this class of inflationary dynamics could in principal be ruled out by future experiments.

This analysis thus reveals that up to a moderate prior dependence, \Pade expansions of the Hubble function give rise to predictions in agreement with observations, and possess good convergence properties. Given what we observationally know, they seem better suited to parametrize inflationary dynamics than Taylor expansions, on which the standard horizon flow procedure rests.
\section{Conclusion}
\label{sec:Conclusion}
Let us now summarize our main findings. Whenever inflation is parametrized by a truncated dynamical system for the flow parameters, it can equivalently be described by an expansion scheme for the Hubble function $H(\phi)$, at some finite order. Conversely, any functional shape for the Hubble function (such as a Taylor expansion, a \Pade expansion, or any other expansion involving a finite set of free coefficients) can be related to a single dynamical system in the flow parameters space.

Making use of the shift symmetry $\phi\to\phi+\delta\phi$ of the problem, we have explained how constants of motion can be derived for such systems, and how their dynamics can be integrated. For illustrative purpose, we have applied this new method to the case of a second order Taylor expansion (section~\ref{sec:analytics:exact:tay}), a first order \Pade expansion (section~\ref{sec:analytics:exact:pad}), and a second order inverse Taylor expansion (appendix~\ref{sec:InverseTaylor}). For the second order Taylor case, we have found that generically, either $r$ is too large and the famous horizon flow relation $r\approx 16 (1-\nS)/3$ is recovered, or $r$ is small enough but $\nS$ is too red. For the first order \Pade expansion on the other hand, the typical predictions $\nS\simeq 1-4/(3\Delta N_*)$ and $r \sim \Delta N_*^{-4/3}\ll 1$ have been obtained, in good agreement with observations. 

We have then extended these results to higher order expansions by numerical means in section~\ref{sec:numeric}, and studied the dependence of the predictions on the priors chosen for the coefficients of the expansions. We have confirmed that \Pade expansions of the Hubble function are more suited to parametrize inflation than Taylor expansions, since they show good convergence properties, mild prior dependence, and, most notably, much better agreement with observations. 

\begin{figure*}[t]
\includegraphics[width=0.5\columnwidth]{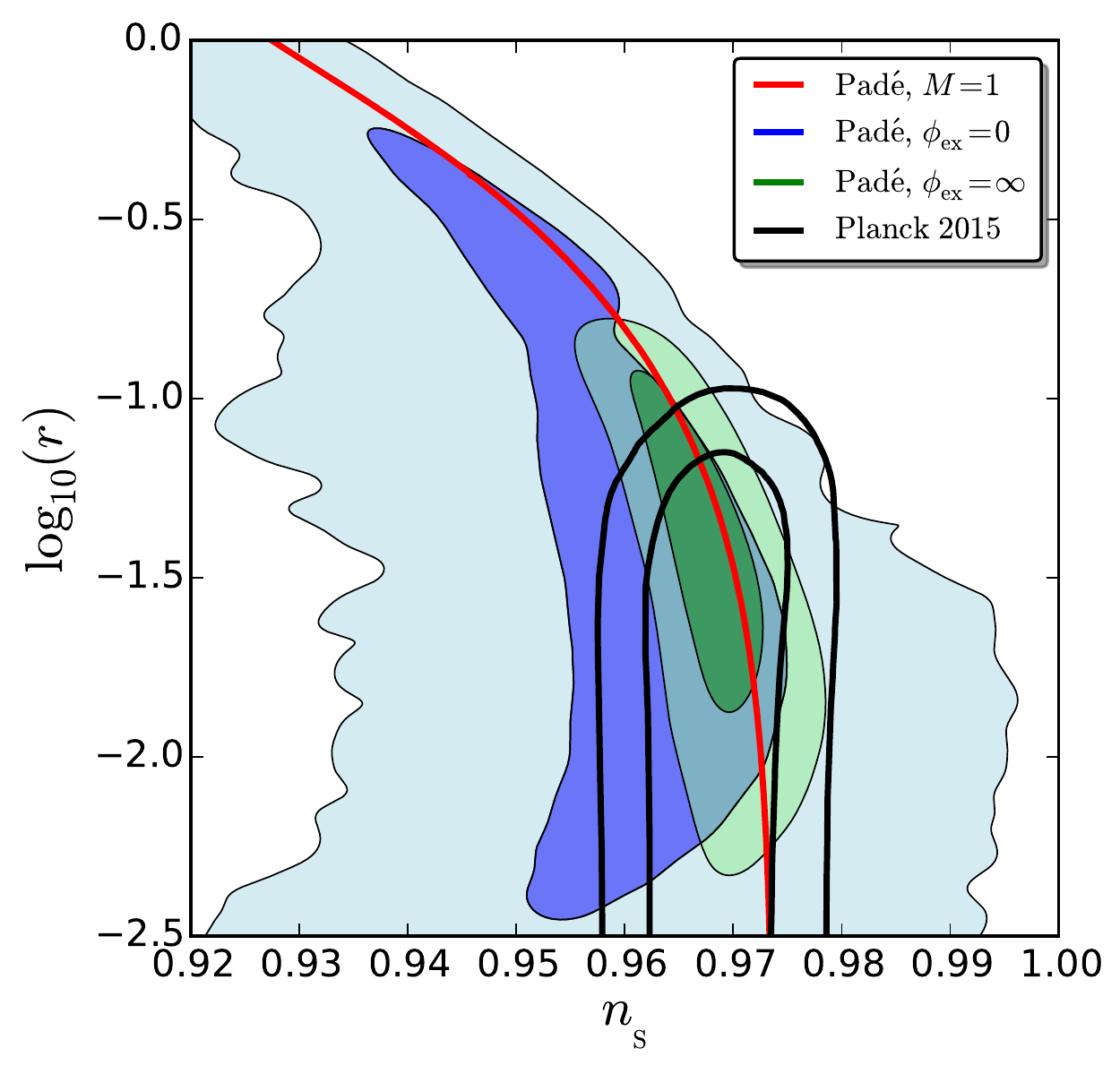}
\includegraphics[width=0.5\columnwidth]{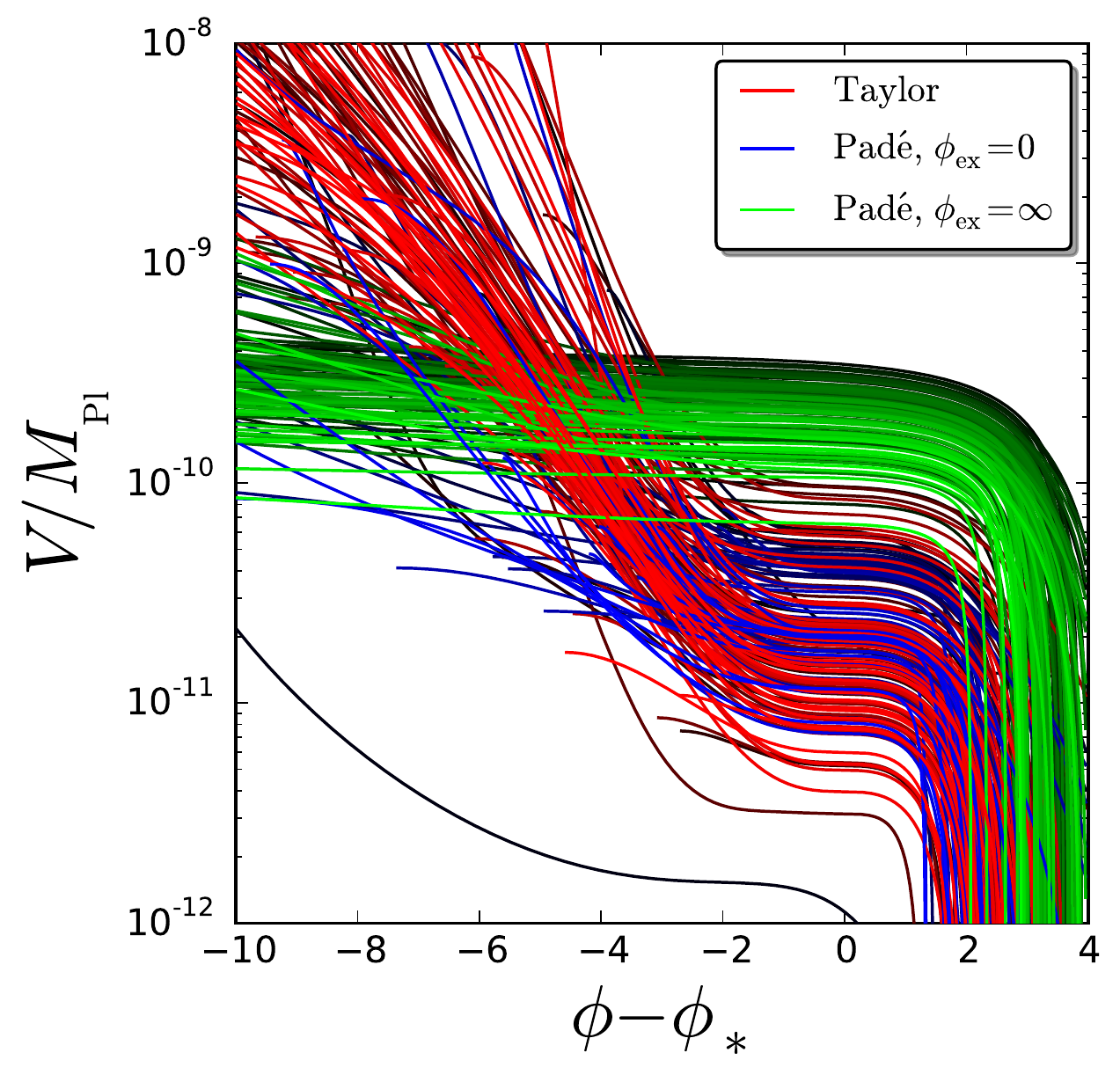}
  \caption{\it Left frame: $1\sigma$ and $2\sigma$ contours for the two \Pade parametrization around $\phi=0$ and $\phi=\infty$, both with power-law priors. The predictions of the first order \Pade expansion, as obtained in section~\ref{sec:analytics:exact:pad}, are displayed too. Right frame: potential reconstruction for the the most likely $100$ trajectories (using the Planck likelihood~\cite{Ade:2015lrj}), for the standard horizon flow procedure based on a Taylor expansion of the Hubble function, a \Pade expansion around $\phi=0$, and a \Pade expansion around $\phi=\infty$. Darker colors stand for smaller likelihood of the predicted values for $\nS$ and $r$. The potentials are normalized so that the amplitude of the curvature power spectrum, $\calP_\zeta=V_*/(24\pi^2\epsilon_{1*}\Mp^4)\simeq 2.203\times 10^{-9}$, is correctly obtained.}
  \label{fig:concl}
\vspace{-0.5cm}
\end{figure*}

When using \Pade approximants, we have distinguished the case where inflation proceeds close to $\phi=0$ and close to $\phi=\infty$. These two prescriptions give rise to results that are compared in the left panel of \Fig{fig:concl}, where we also display the first order \Pade result of section~\ref{sec:analytics:exact:pad}. It is clear that, given observational constraints, they provide a better parametrization of inflationary dynamics than the usual horizon flow procedure.

These results illustrate why ``model-independent'' parametrizations of inflation, such as expansion schemes for the Hubble function or truncated flow dynamical systems, always make non trivial assumptions about its dynamics. This is why, a priori, parametrizations yielding predictions that are in contradiction with the data cannot be used to infer physical information from them.

As an example, in the right panel of \Fig{fig:concl}, we show the best 100 potentials, obtained from the formula $V=3\Mp^2 H^2-2\Mp^4 {H^\prime}^2$, for the three cases: Taylor expansion with the priors corresponding to the usual ``horizon flow'' procedure, \Pade expansion around $\phi=0$, and \Pade expansion around $\phi=\infty$. From here, it is clear that different parametrizations sample inflation along different classes of potentials, even when restricted to the best possible realizations. Taylor expansion and \Pade expansions around $\phi=0$ seem to prefer inflationary potentials with a flat inflection point. Because this is a rather fine-tuned configuration from a generic Taylor expansion, we understand why the standard horizon flow procedure, which is based on a generic Taylor expansion, produces trajectories that are most of the time excluded by observations. On the contrary, \Pade expansions around $\phi=\infty$ mostly samples plateau inflation, as expected.

As a consequence, it is clear that ``reconstructing'' the potential with either of these parametrizations biases the result towards the class of potentials that it relies on. This is in essence a ``prior effect''. 
However, phenomenological descriptions are still very useful to address a number of other issues in the Early Universe, where the background is effectively ``sourcing'' some physical effects. Therefore, the question becomes: ``How can we best obtain and parametrize a class of inflationary trajectories that are in agreement with current observational constraints?'' From a Bayesian perspective, the priors for analyzing the $n{}^\mathrm{th}$ generation of data come from the information provided by the $n-1{}^\mathrm{th}$ survey. In this respect, we have shown that after Planck, \Pade expansions (or other types of expansion schemes implementing the plateau structure) should be preferred over the standard Taylor parametrizations of the Hubble flow dynamics.
\section*{Acknowledgments}
We acknowledge the Millipede Cluster at the University of Groningen, that we used for the numerical simulations. VV's work is supported by STFC grant ST/L005573/1.
\appendix
\section{Inverse Taylor Expansion}
\label{sec:InverseTaylor}
In order to further illustrate the method depicted in section~\ref{sec:FlowDynamics}, in this appendix, we apply it to the case where the inverse Hubble function is Taylor expanded at second order,
\beq
\label{eq:InverseTaylorHubble}
\frac{H}{H_0}=\frac{1}{1+a\phi+b\phi^2}\, .
\eeq
The two first slow-roll parameters can be read off from \Eqs{eq:eps:H}, and one has
\beq
  \epsilon_1 = \frac{2\left(a+2b\phi\right)^2}{\left(1+a\phi+b\phi^2\right)^2} \,, \quad \epsilon_2 =-4\frac{2b^2\phi^2+2ab\phi+a^2-2b}{\left(1+a\phi+b\phi^2\right)^2}\, .
\eeq
Then, under shift transformations $\phi\rightarrow\phi+\delta\phi$, the functional form~(\ref{eq:InverseTaylorHubble}) is unchanged provided
\beq
  	a \to \frac{a+2b\delta\phi}{1+a\delta\phi+b\delta\phi^2} \,, \quad 
	b \to \frac{b}{1+a\delta\phi+b\delta\phi^2} \,,
\eeq
where $H_0$ is also rescaled according to $H_0 \to H_0/(1+a\delta\phi+b\delta\phi^2)$. These gauge transformations are, for obvious reasons, the same as for the second order Taylor case studied in section~\ref{sec:analytics:exact:tay}, which implies that
\beq
\label{eq:def:gamma:InverseTaylor}
	\gamma=\frac{32b^2}{4b-a^2} = \frac{\left(2\epsilon_1+\epsilon_2\right)^2}{\epsilon_1+\epsilon_2}
\eeq
is a constant of motion and can be used to label the different trajectories.

\begin{figure*}[t]
\begin{center}
\includegraphics[width=\widthdouble]{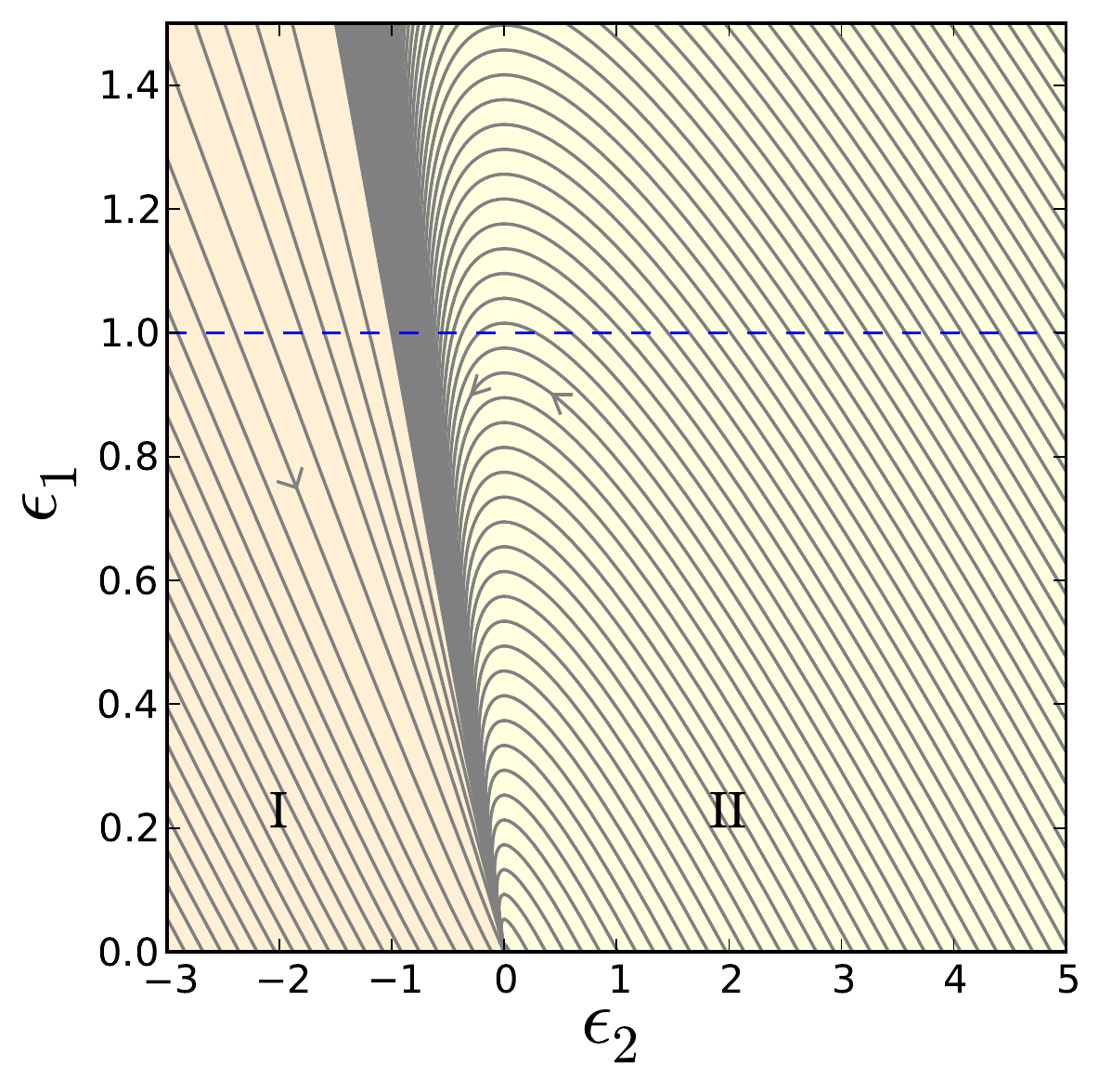}
\caption{\it Inverse second order Taylor expansion for the Hubble function: flow lines of the system~(\ref{eq:depsdN:InverseTaylor}) in the plane $(\epsilon_1,\epsilon_2)$. The arrows indicate in which direction inflation proceeds. The blue dashed line corresponds to $\epsilon_1=1$ where inflation stops. The two regions $\rm I$ and $\rm II$ refer to the discussion in the main text.}
\label{fig:InverseTaylor}
\end{center}
\vspace{-0.5cm}
\end{figure*}

Let us now derive the dynamical system associated to this parametrization of the Hubble function. Making use of the same procedure as in section~\ref{sec:analytics}, we find that \Eq{eq:InverseTaylorHubble} implies that $H^{\prime\prime\prime}=6HH^\prime H^{\prime\prime}/H^2$, and \Eqs{eq:eps:H} then gives rise to
\beq
\label{eq:eps3:InverseTaylor}
   \epsilon_3 = -\epsilon_1\left(3+2\frac{\epsilon_1}{\epsilon_2}\right) \,.
\eeq
This truncates the dynamical system to a closed set of differential equations for $(\epsilon_1, \epsilon_2)$, given by
\beq
\label{eq:depsdN:InverseTaylor}
   \frac{\dd \epsilon_1}{\dd N} = \epsilon_1 \epsilon_2 \,, \qquad
   \frac{\dd \epsilon_2}{\dd N} = -\epsilon_1\left(3\epsilon_2+2\epsilon_1\right) \,. 
\eeq
In particular, one can check that the combination $\gamma$ defined in \Eq{eq:def:gamma:InverseTaylor} is left invariant. The integrated flow lines of the above system are displayed in \Fig{fig:InverseTaylor}. 

Let us now see how this system can be integrated analytically. By inverting \Eq{eq:def:gamma:InverseTaylor}, one can express $\epsilon_2$ as a function of $\epsilon_1$,
\beq 
\label{eq:eps2ofeps1:InverseTaylor}
   \epsilon_2 = -2\epsilon_1+\frac{\gamma}{2}+\frac{\xi}{2}\sqrt{\gamma^2-4\gamma\epsilon_1} \,,
\eeq
where\footnote{This sign does not change for trajectories for which $\epsilon_1$ always decreases with time (regions I in the classification introduced below). However, for trajectories for which $\epsilon_1$ first increases, reaches a maximum and then decreases, it is positive before reaching the maximum and negative afterwards (region II).} $\xi=\pm 1 =\mathrm{sign}\left[\frac{\epsilon_2(2\epsilon_1+\epsilon_2)}{\epsilon_1+\epsilon_2}\right]$. As before, inserting \Eq{eq:eps2ofeps1:InverseTaylor} into \Eqs{eq:depsdN:InverseTaylor} yields a first order differential equation for $\epsilon_1(N)$ that can be solved, and one obtains $\Delta N_*=N(\epsilon_{1,\mathrm{end}})-N(\epsilon_{1*})$, where $\epsilon_{1,\mathrm{end}}=1$ and
\beq
\label{eq:InverseTaylor:nef:2}
 N \left(\epsilon_1\right) = \frac{2}{\xi\sqrt{\gamma^2-4\gamma\epsilon_1}-\gamma}+\frac{\xi}{2\gamma}\log \left\vert\frac{\sqrt{\gamma^2-4\gamma\epsilon_1}+\gamma}{\sqrt{\gamma^2-4\gamma\epsilon_1}-\gamma}\right\vert \, .
\eeq
As in the case of the Taylor parametrization, \Eq{eq:InverseTaylor:nef:2} can be inverted,
\beq
\label{eq:InverseTaylor:eps1N}
 \epsilon_1(N)=\frac{-4 W_\chi\left(-e^{-\gamma\Delta N-1}\right)}{\left[ 1+W_\chi\left(-e^{-\gamma\Delta N-1}\right) \right]^2} \,, \quad \chi = \begin{cases} 
				0 \text{ if } \gamma(2\epsilon_1+\epsilon_2)>0 \\
				-1 \text{ if } \gamma(2\epsilon_1+\epsilon_2)<0
\end{cases}\, ,
\eeq
where $\chi$ determines the branch of the Lambert function $W_\chi$. It is easy to show that $\chi$ does not change along a given trajectory.

Let us now discuss the structure of the phase space diagram plotted in \Fig{fig:InverseTaylor}. According to the type of Hubble function one is dealing with, two possibilities must be distinguished:\\

\begin{minipage}{0.7\textwidth}
\begin{itemize}
\item
Firstly, if $\epsilon_2 <- \epsilon_1$, $\epsilon_1$ decreases and $\epsilon_2$ increases as inflation proceeds, reaching one of the (attractive) fixed point $(\epsilon_1 =0, \epsilon_2<0)$ in the asymptotic future. The corresponding Hubble function has a convex shape with a positive minimum (region I). In this case, $\gamma<0$.\\
\end{itemize}
\end{minipage} \hfill
\begin{minipage}[c]{0.2\textwidth}
\begin{center}
\includegraphics[width=\textwidth]{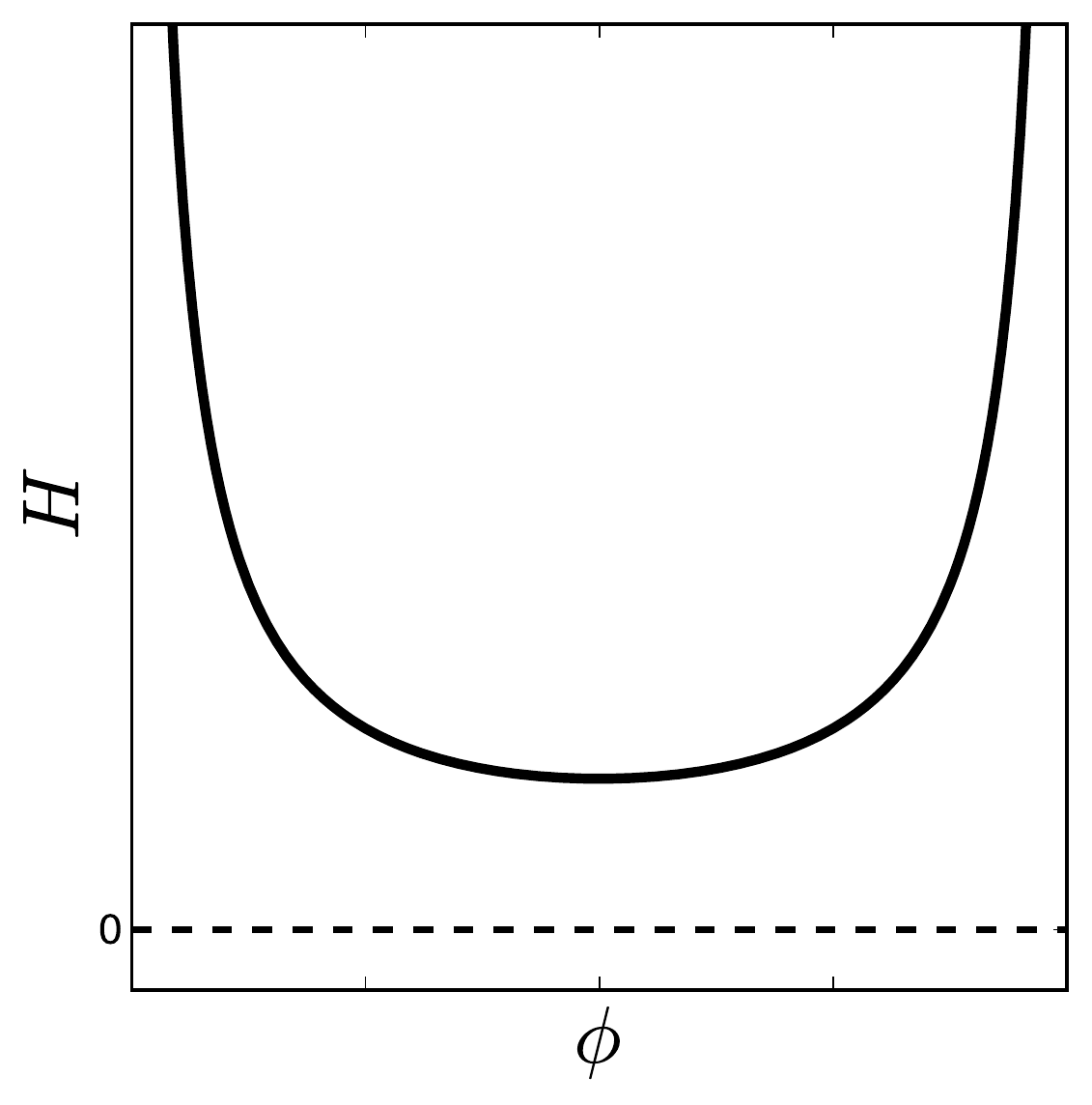}
\end{center}
\end{minipage}

\begin{minipage}{0.7\textwidth}
\begin{itemize}
\item
Secondly, if $\epsilon_2 > -\epsilon_1$, $\epsilon_2$ decreases as inflation proceeds while $\epsilon_1$ first increases, crosses a maximum and then decreases, reaching the (attractive) fixed point $(\epsilon_1 =0, \epsilon_2=0)$ in the asymptotic future. The corresponding Hubble function is concave and vanishes at infinity (region II). In this case, $\gamma>0$.\\
\end{itemize}
\end{minipage} \hfill
\begin{minipage}[c]{0.2\textwidth}
\begin{center}
\includegraphics[width=\textwidth]{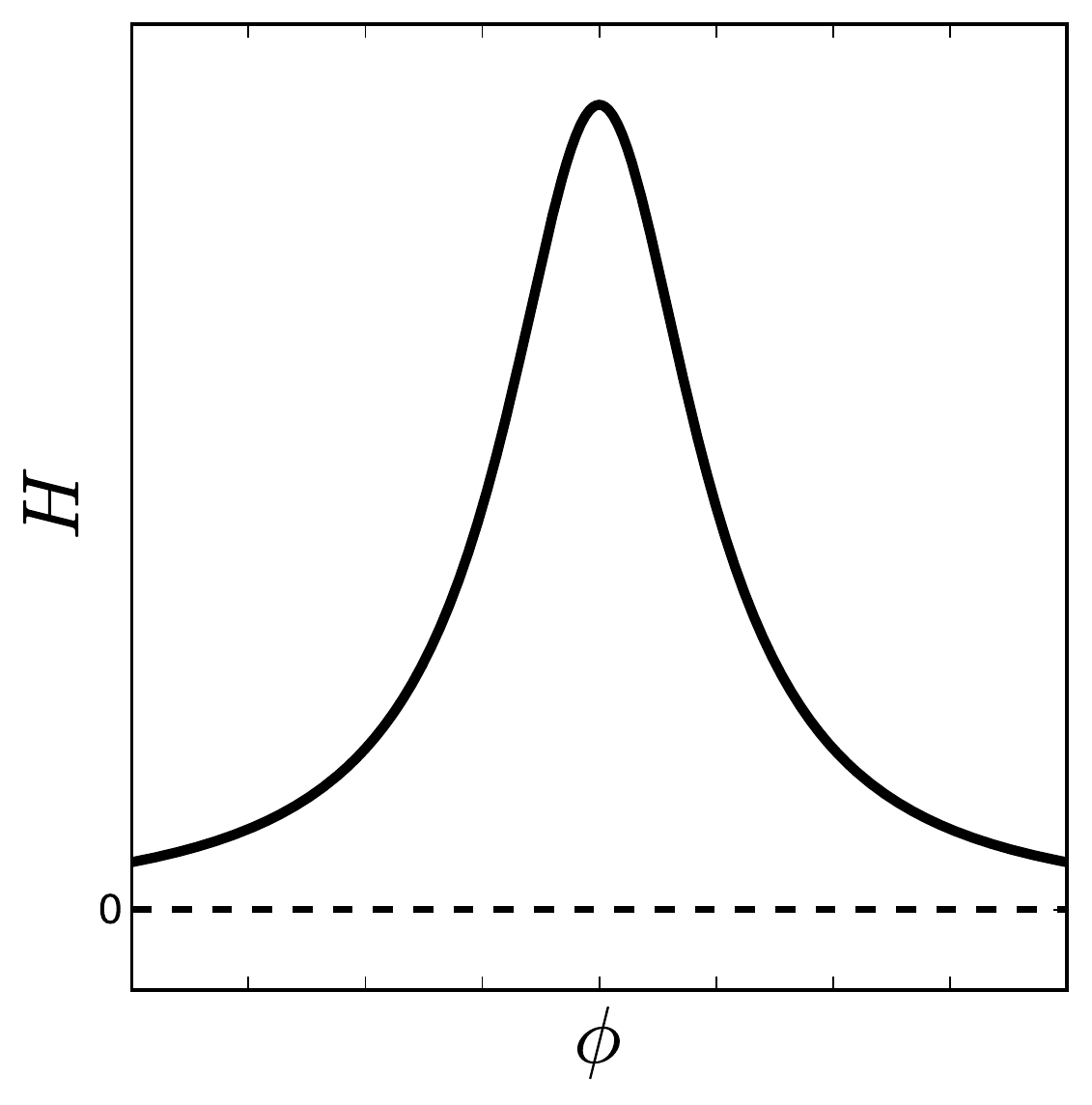}
\end{center}
\end{minipage}
One should note that thanks to the conservation of the sign of $\gamma$ defined in \Eq{eq:def:gamma:InverseTaylor}, a given inflationary trajectory never changes regions. Amongst the second category, one can distinguish two cases. If the maximum value of $\epsilon_1$ is smaller than one, inflation never ends and reaches the fixed point $(\epsilon_1 =0, \epsilon_2=0)$. If, on the other hand, the maximum value of $\epsilon_1$ is larger than one, and if one starts inflating with $\epsilon_2>0$, then inflation ends naturally when $\epsilon_1=1$. This happens when $\gamma>4$. In this case, one can check that the function $N \left(\epsilon_1\right)$ defined in \Eq{eq:InverseTaylor:nef:2} goes to infinity when $\epsilon_1$ goes to $0$ which
means that a sufficient number of e-folds can always be realized.

However, this requires $\epsilon_{1*}$ to be sufficiently small. If one parametrizes a given trajectory within region II by $\epsilon_{2,\uend}=-2+(\gamma+\sqrt{\gamma^2-4\gamma})/2$, the value of the second flow parameter at the end of inflation, in the $\epsilon_{1*}\ll 1$ limit one  has
\beq
\epsilon_{1*}\simeq \left(\frac{2+\epsilon_{2\uend}}{1+\epsilon_{2\uend}}\right)^2\exp\left[\frac{1-\left(2+\epsilon_{2\uend}\right)^2\Delta N_*}{1+\epsilon_{2\uend}}\right]\, .
\eeq
Since $\epsilon_{2*}>\epsilon_{2\uend}$, $\epsilon_{2*}<1$ implies that $\epsilon_{2\uend}<1$ hence $\epsilon_{1*}< 10^{-104}$ if one lets $\Delta N_*=60$. So essentially, $r\simeq 0$ in these models. Making use of \Eq{eq:eps2ofeps1:InverseTaylor}, one then has
\beq
\epsilon_{2*}\simeq \frac{\left(2+\epsilon_{2\uend}\right)^2}{1+\epsilon_{2\uend}}\, .
\eeq
Because $\epsilon_{2\uend}>0$ in those branches where inflation ends naturally, this means that $\epsilon_{2*}>4$, which is of course completely excluded by CMB observations. As a conclusion, the only trajectories compatible with observations are those that reach the fixed points $(\epsilon_1=0,\epsilon_2<0)$ such that $\epsilon_2$ gives the correct value of $\nS$. However, these are rather fine-tuned situations that moreover require to invoke an extra mechanism to end inflation.

\bibliographystyle{JHEP}
\bibliography{plateau}
\end{document}